\documentclass[smallextended]{svjour3}

\usepackage[parfill]{parskip}  
\usepackage{amsmath,amssymb,amsfonts,amsbsy,amsfonts,latexsym}
\usepackage{multirow}
\usepackage{makecell}
\usepackage[labelfont=bf,belowskip=0pt,aboveskip=5pt,tableposition=top, font=footnotesize]{caption}
\usepackage{xcolor}
\usepackage{colortbl}
\usepackage[sort]{natbib}

\usepackage{graphicx}     
\DeclareGraphicsExtensions{.pdf,.png}

\usepackage{subcaption}

\usepackage{float}
\usepackage{mathtools}

\newcommand{\secref}[1]{\S\ref{#1}}

\newcommand{\vect}[1]{\mathbf{#1}} 

\definecolor{light-gray}{gray}{0.80}

\renewcommand\paragraph{\subsubsection*}

\usepackage[colorlinks=true,
citecolor=blue,
filecolor=black,
linkcolor=blue,
urlcolor=blue]{hyperref}

\newcommand\hf{ \rowcolor{gray!30}}
\newcommand\ghligh{ \hf}

\newcommand\eref{Eq. \ref}

\newcommand\fref{Fig. \ref}
\newcommand\tref{Table \ref}

\definecolor{colorA}{RGB}{189,201,225}
\definecolor{colorB}{RGB}{103,169,207}
\definecolor{colorC}{RGB}{ 28,144,153}
\definecolor{colorD}{RGB}{  1,108, 89}

\newcommand{\vI}{\vect{I}}

\newcommand{\vomega}{\vect{\omega} }

\newcommand{\vu}{\vect{u}}

\newcommand{\vE}{\vect{E}}
\newcommand{\vX}{\vect{X}}
\newcommand{\vzero}{\vect{0}}
\newcommand{\vx}{\vect{x}}

\newcommand{\vp}{\vect{p}}
\newcommand{\vK}{\vect{K}}
\newcommand{\R}{\rm I\!R }

\renewcommand{\div}{{\rm div}\ }
\renewcommand{\div}{{\nabla \cdot}\ }

\newcommand{\igrad}{\ensuremath{\nabla}}
\newcommand{\ilap}{\rotatebox[origin=c]{180}{$\nabla$}}

\journalname{J. Math. Biol.}	
\begin{document}
\title{Simulation of glioblastoma growth using a 3D multispecies tumor model with mass effect}   
\titlerunning{3D multispecies tumor model with mass effect}                          
\author{Shashank Subramanian \and Amir Gholami \and George Biros}
\authorrunning{S. Subramanian et al.}
\institute{S.Subramanian \and G.Biros \at 
	Institute for Computational Engineering and Sciences, University of Texas at Austin, TX 78712, USA. \\
	\email{shashanksubramanian@utexas.edu, biros@ices.utexas.edu}
	\and
	A.Gholami \at
	Department of Electrical Engineering and Computer Sciences, UC Berkeley, CA 94720, USA. \\
	\email{amirgh@berkeley.edu }
}
\maketitle
\begin{abstract}

In this article, we present a multispecies reaction-advection-diffusion partial
differential equation (PDE) coupled with linear elasticity for modeling tumor
growth.  The model aims to capture the phenomenological features of
glioblastoma multiforme observed in
magnetic resonance imaging (MRI) scans. These include enhancing and necrotic
tumor structures, brain edema and the so called \emph{``mass effect''}, that
is, the deformation of brain tissue due to the presence of the tumor. The
multispecies model accounts for proliferating, invasive and necrotic tumor
cells as well as a simple model for nutrition consumption and tumor-induced
brain edema. The coupling of the model with linear elasticity equations with
variable coefficients allows us to capture the mechanical deformations due to
the tumor growth on surrounding tissues. We
present the overall formulation along with a novel operator-splitting scheme
with components that include linearly-implicit preconditioned elliptic solvers,
and semi-Lagrangian method for advection. Also, we present results showing
simulated MRI images which highlight the capability of our method to capture
the overall structure of glioblastomas in MRIs.

\end{abstract}

\section{Introduction} \label{s:intro} Glioblastomas form a class of highly aggressive tumors, accounting for a
majority of all malignant primary brain tumors in adults~\citep{dolecek2012cbtrus}, with a median survival
rate of less than a year~\citep{salcman1980survival, newton1994primary, seither1995results, wrensch2002epidemiology,
	swanson2008mathematical}.

Mathematical modeling of glioblastoma tumor growth has been extensively used to assist in image analysis of MRIs~\citep{gooya2011deformable}, as well as the diagnosis, treatment, and prognosis of glioblastomas~\citep{akbari-e16,macyszyn-e16,hawkins2013modeling,szeto2009quantitative}. Beyond glioblastomas, there is a large body of work for generic tumor growth modeling at different scales and different scenarios (e.g., in vitro, animal models, multiscale models) that attempt to capture the complex biological principles underlying tumor dynamics, by accounting for phenomena on cellular scales or continuum/tissue scales. In this paper, we are primarily interested in capturing the phenomenological features of malignant glioblastomas or glioblastomas, observed from MRI scans. These include

\begin{itemize}
	\itemsep-0.5em 
	\item enhancing rim of proliferating tumor cells,
	\item central tumor core filled with necrotic/dead cells,
	\item brain edema, and
	\item mass effect.
\end{itemize}

Our end goal is to couple this model with parameter estimation methods and with patient MRIs in order to assist in diagnosis and prognosis. The model also finds applications in areas like MR image segmentation of glioblastomas. For example, our model is incorporated in the training of neural networks on synthetic datasets as a data augmentation strategy in~\cite{brats18}. Here we only present the overall formulation. 

\begin{figure}
	\begin{subfigure}{0.2\linewidth}
		\centering
		\includegraphics[height=1\linewidth, width=1\linewidth]{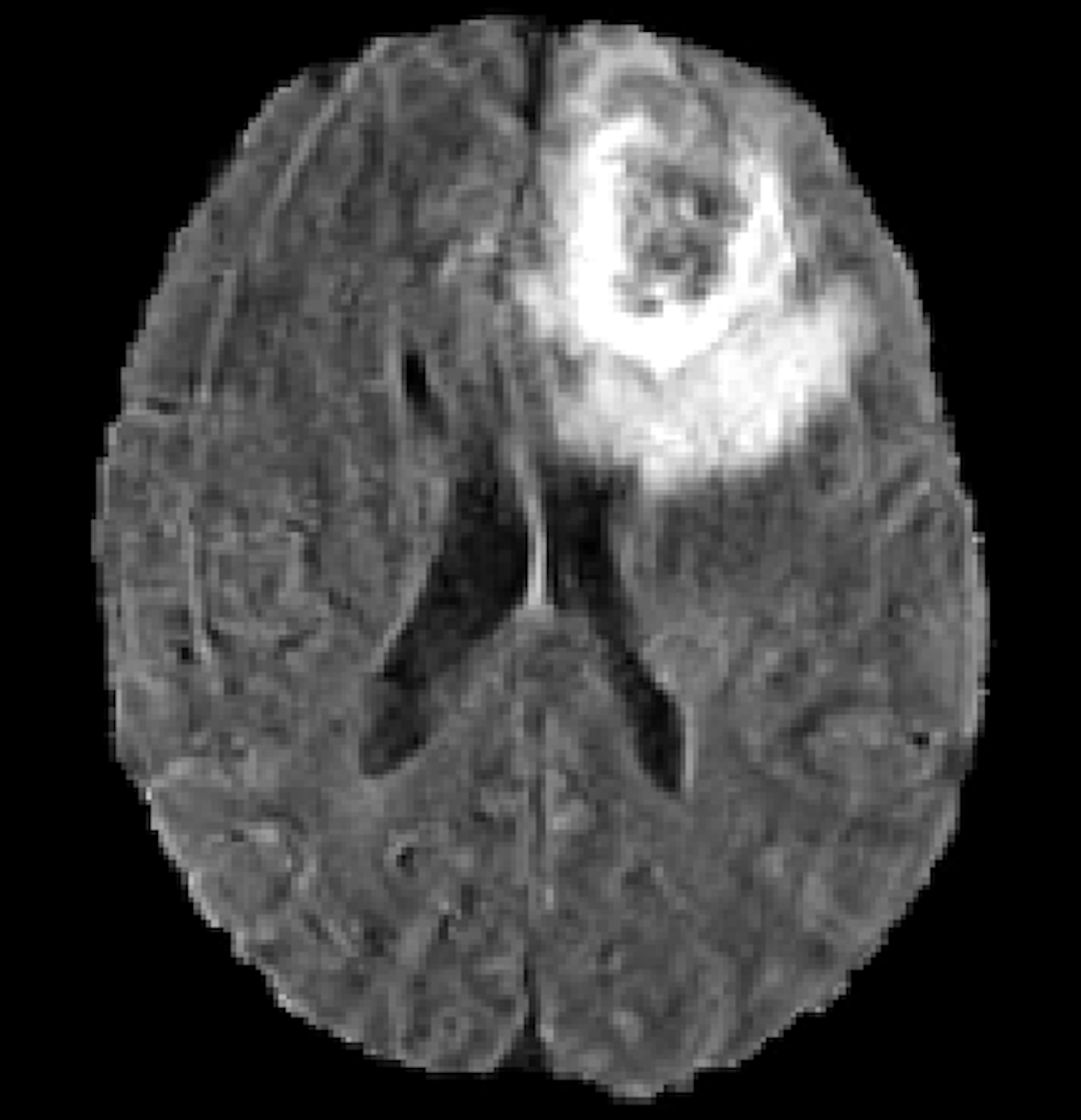}
	\end{subfigure}%
	\begin{subfigure}{0.2\linewidth}
		\centering
		\includegraphics[height=1\linewidth, width=1\linewidth]{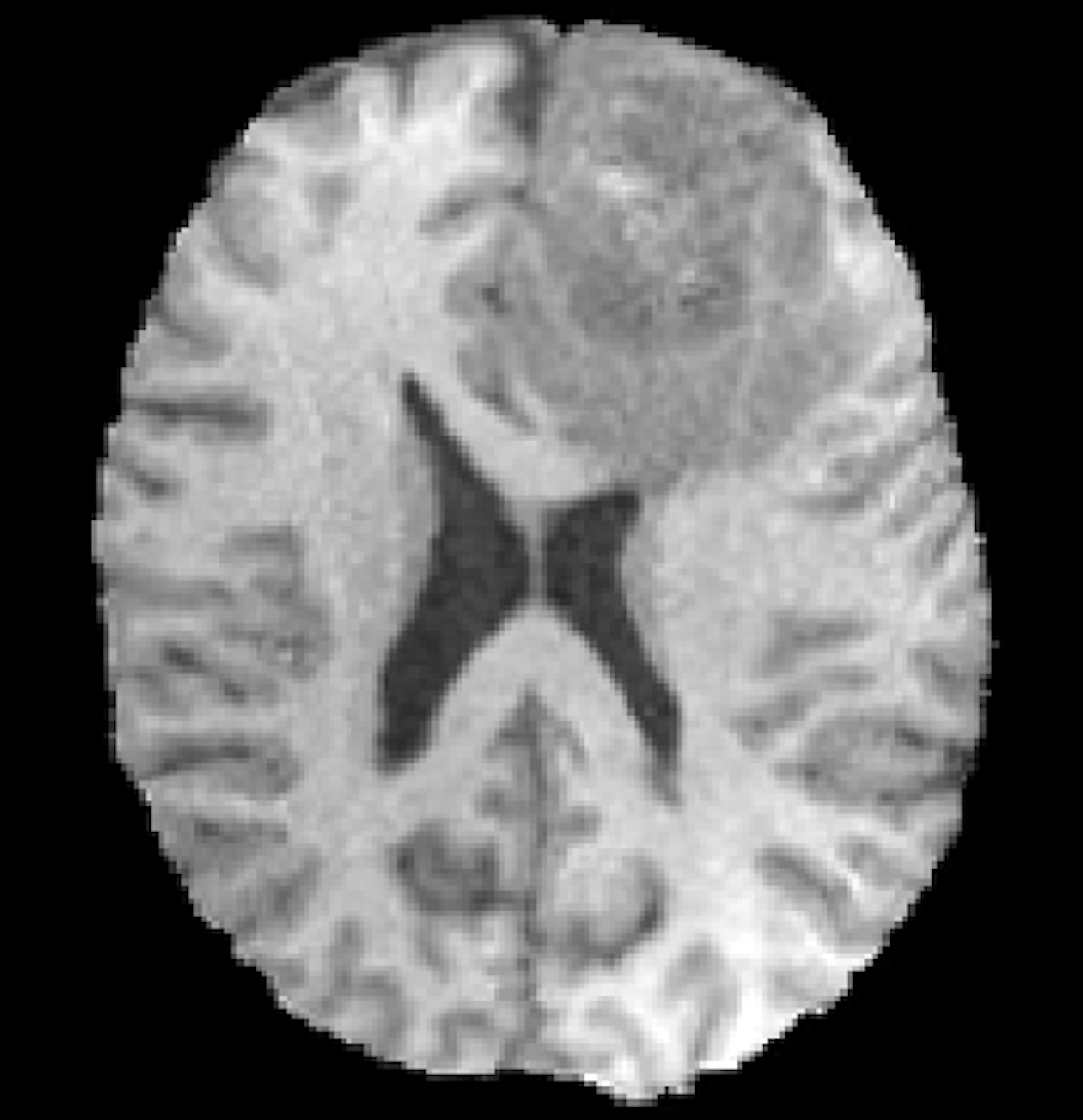}
	\end{subfigure}%
	\begin{subfigure}{0.2\linewidth}
		\centering
		\includegraphics[height=1\linewidth, width=1\linewidth]{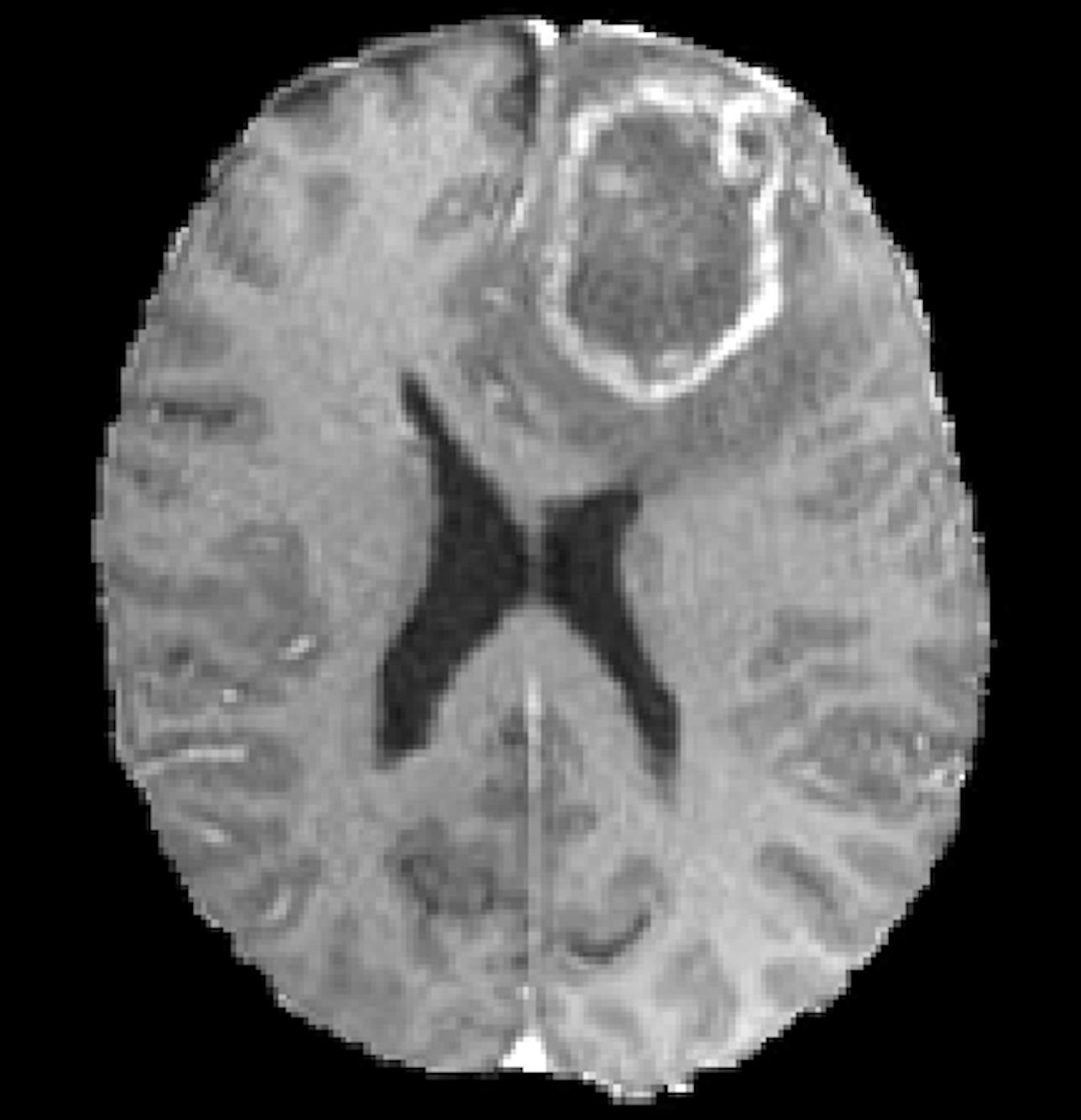}
	\end{subfigure}%
	\begin{subfigure}{0.2\linewidth}
		\centering
		\includegraphics[height=1\linewidth, width=1\linewidth]{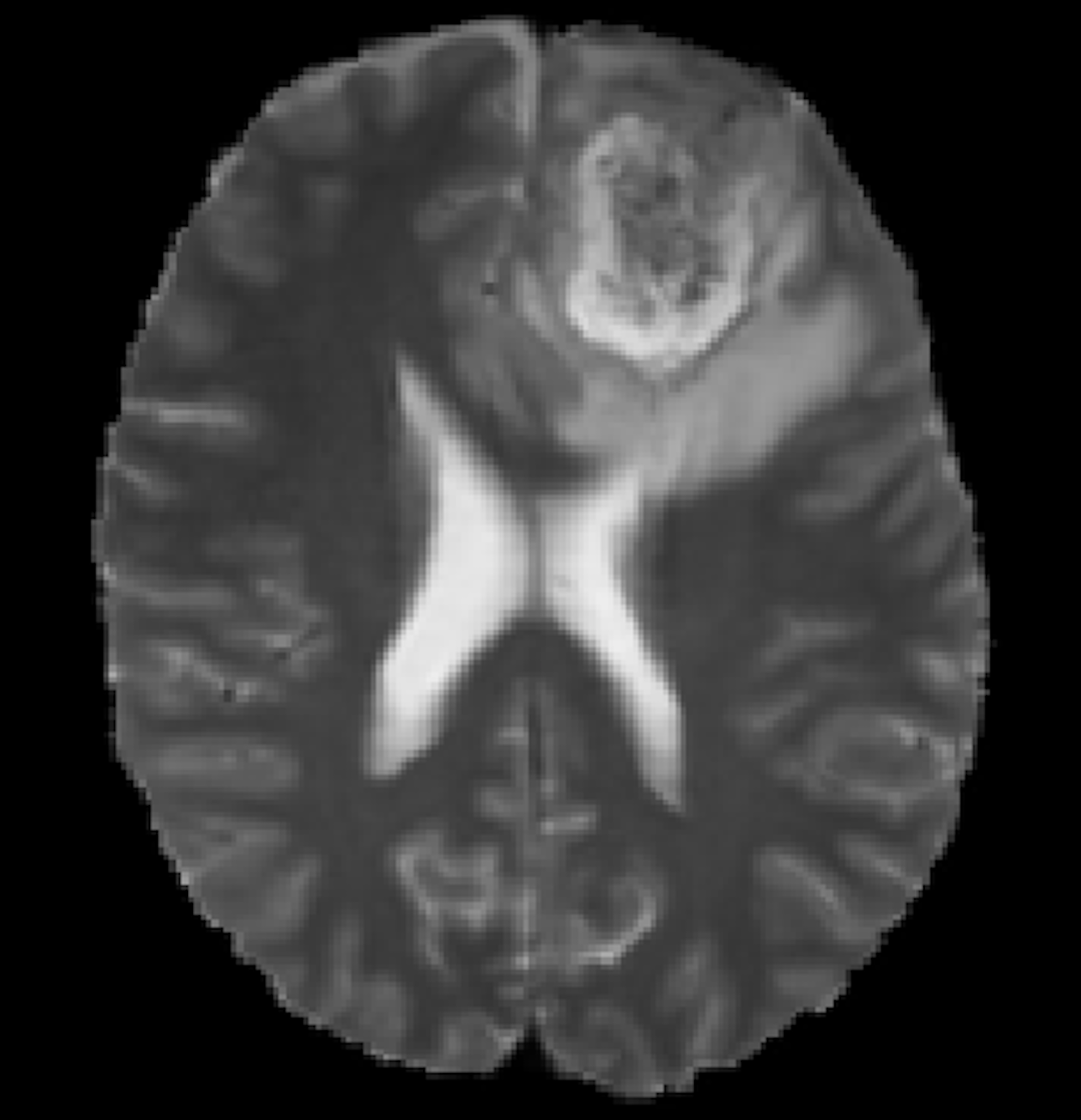}
	\end{subfigure}%
	\begin{subfigure}{0.2\linewidth}
		\centering
		\includegraphics[height=1\linewidth, width=1\linewidth]{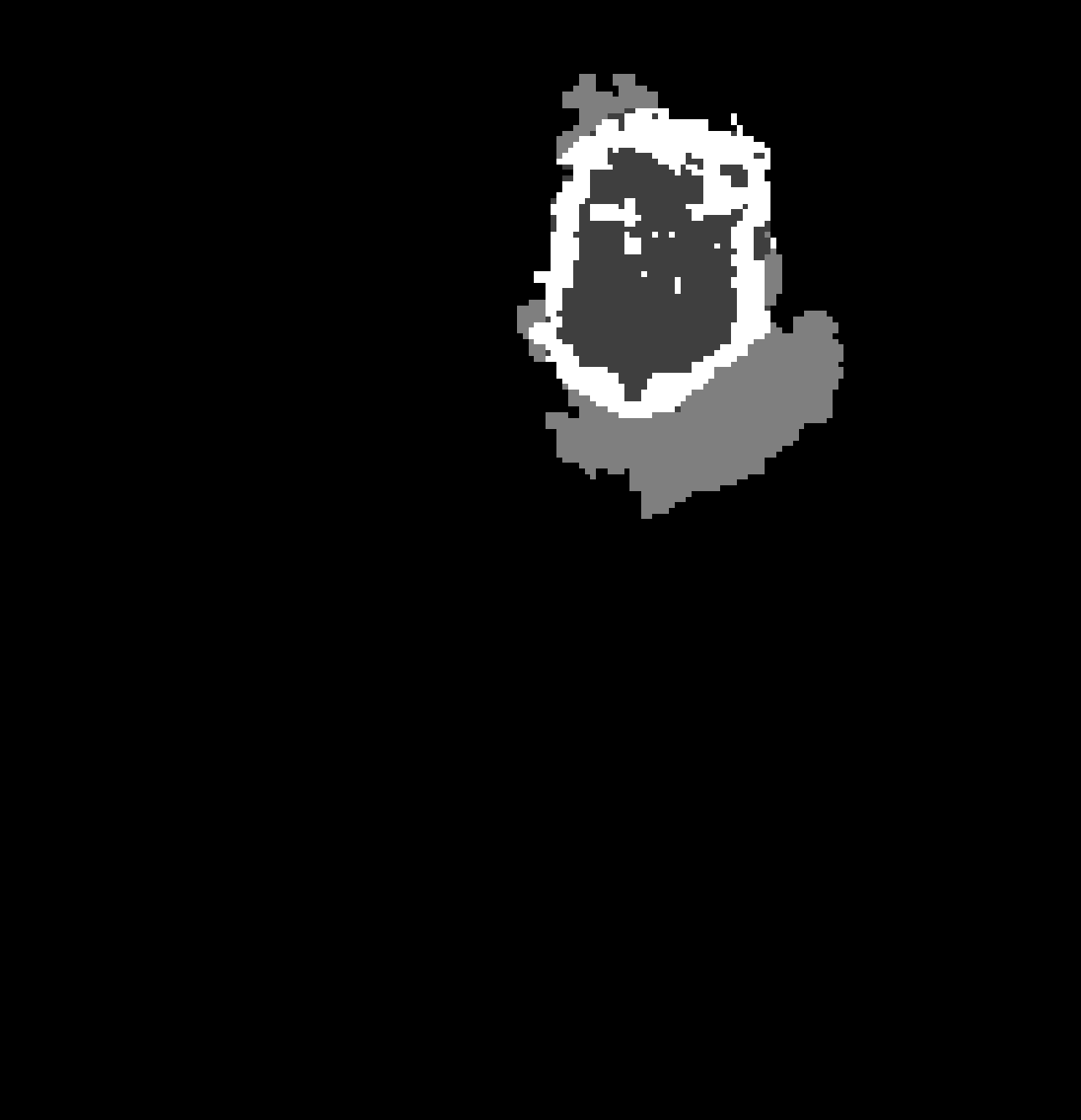}
	\end{subfigure}
	\begin{subfigure}{0.2\linewidth}
		\centering
		\includegraphics[height=1\linewidth, width=1\linewidth]{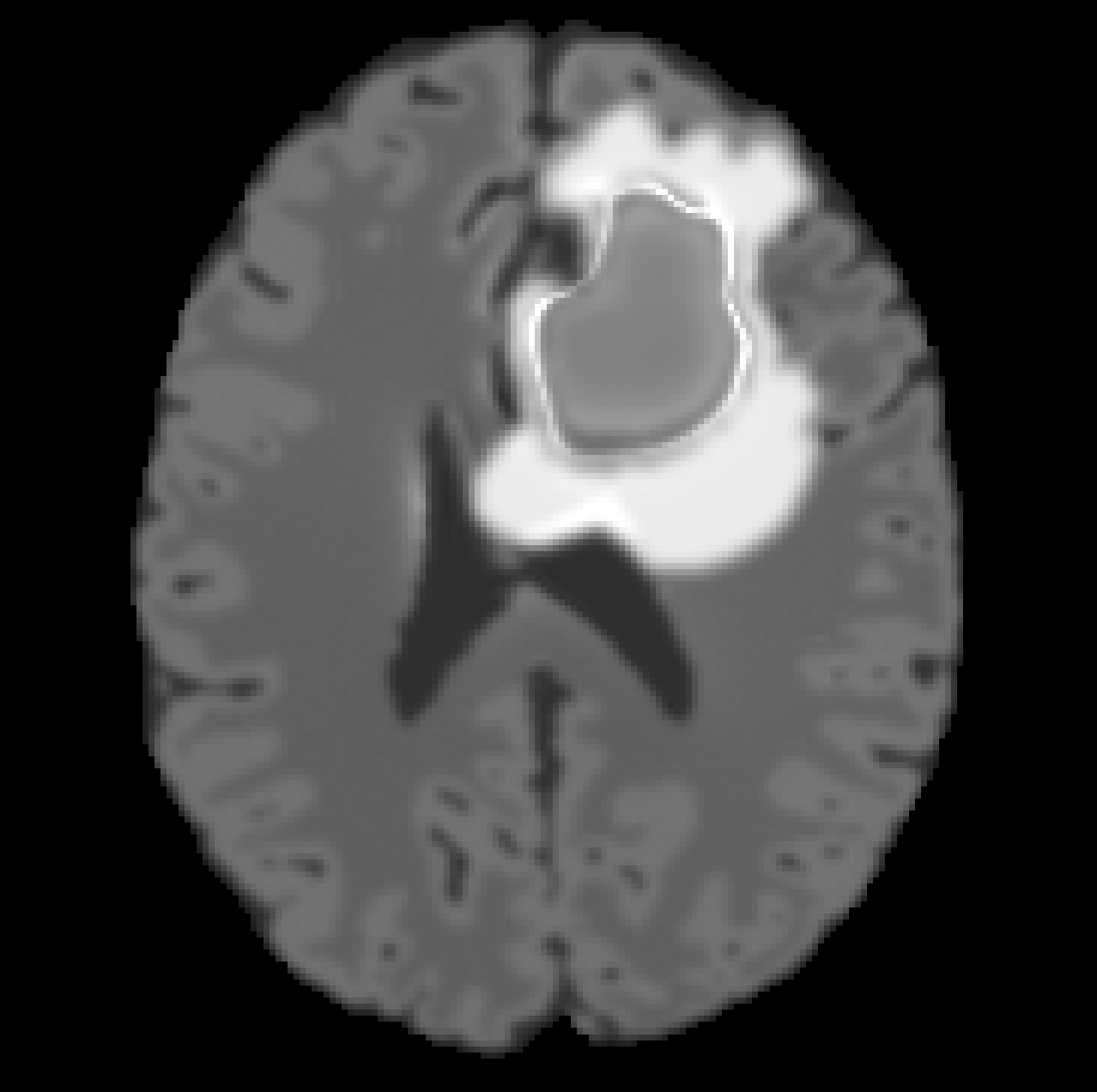}
		\caption{}
	\end{subfigure}%
	\begin{subfigure}{0.2\linewidth}
		\centering
		\includegraphics[height=1\linewidth, width=1\linewidth]{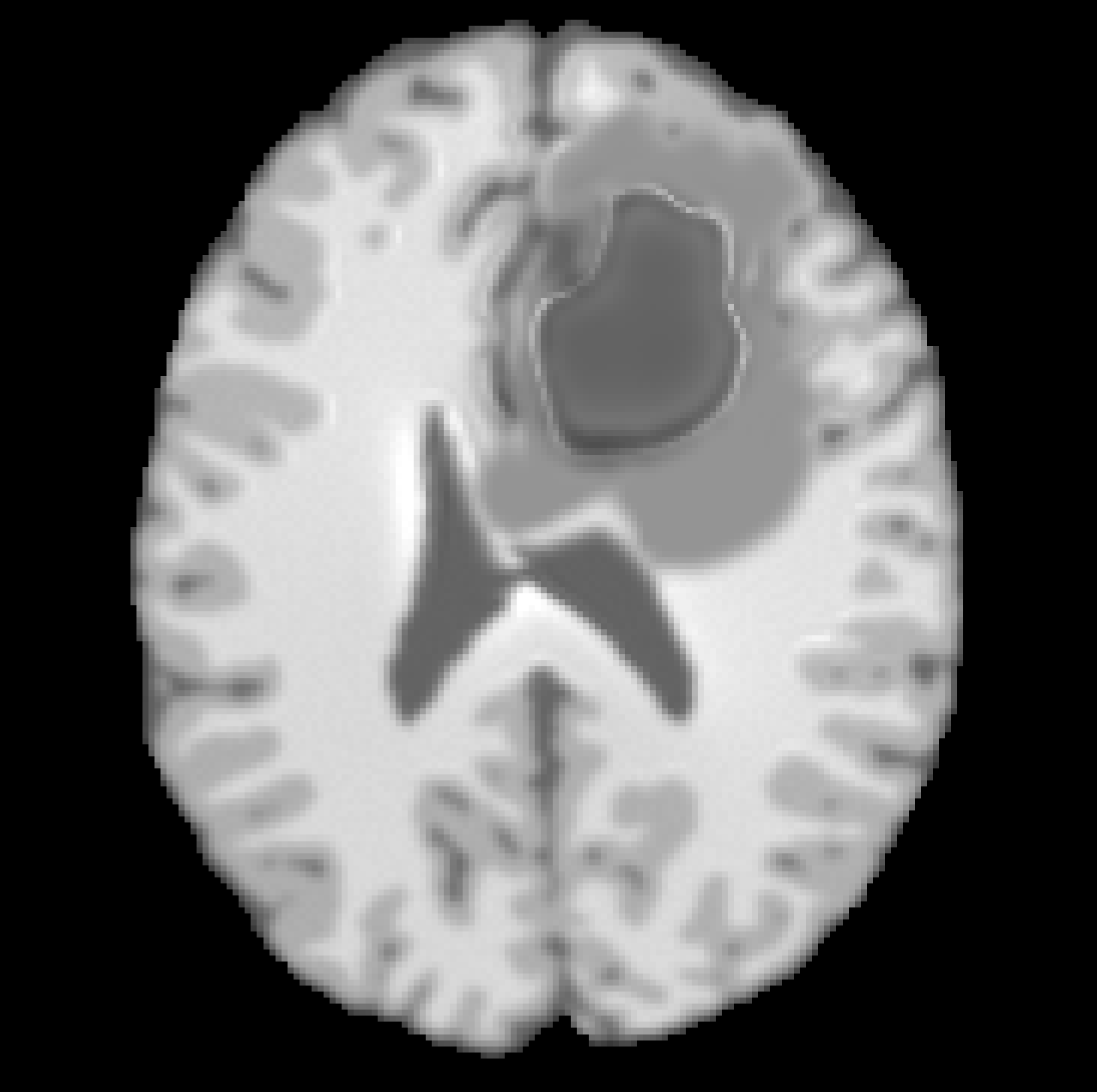}
		\caption{}
	\end{subfigure}%
	\begin{subfigure}{0.2\linewidth}
		\centering
		\includegraphics[height=1\linewidth, width=1\linewidth]{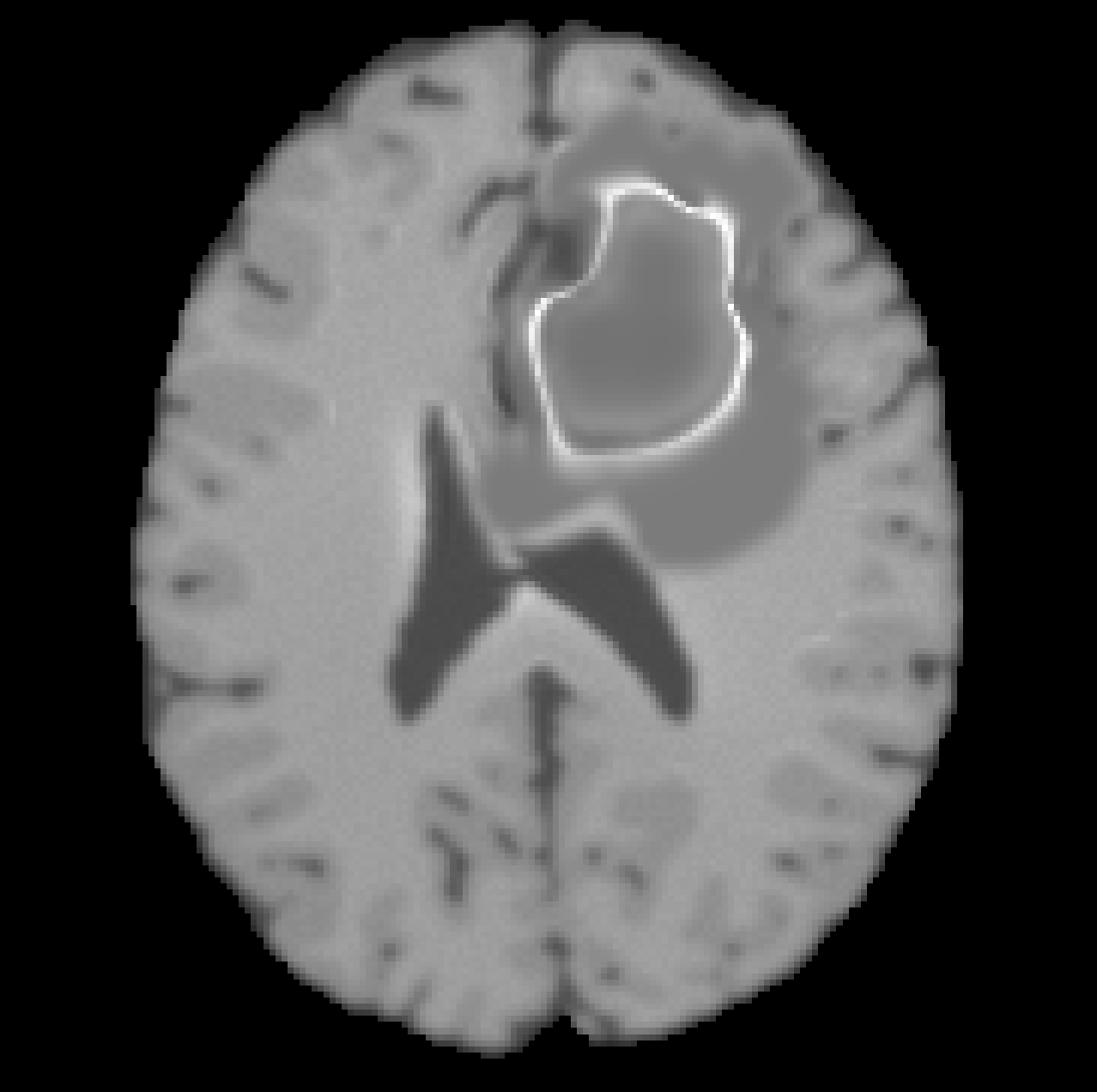}
		\caption{}
	\end{subfigure}%
	\begin{subfigure}{0.2\linewidth}
		\centering
		\includegraphics[height=1\linewidth, width=1\linewidth]{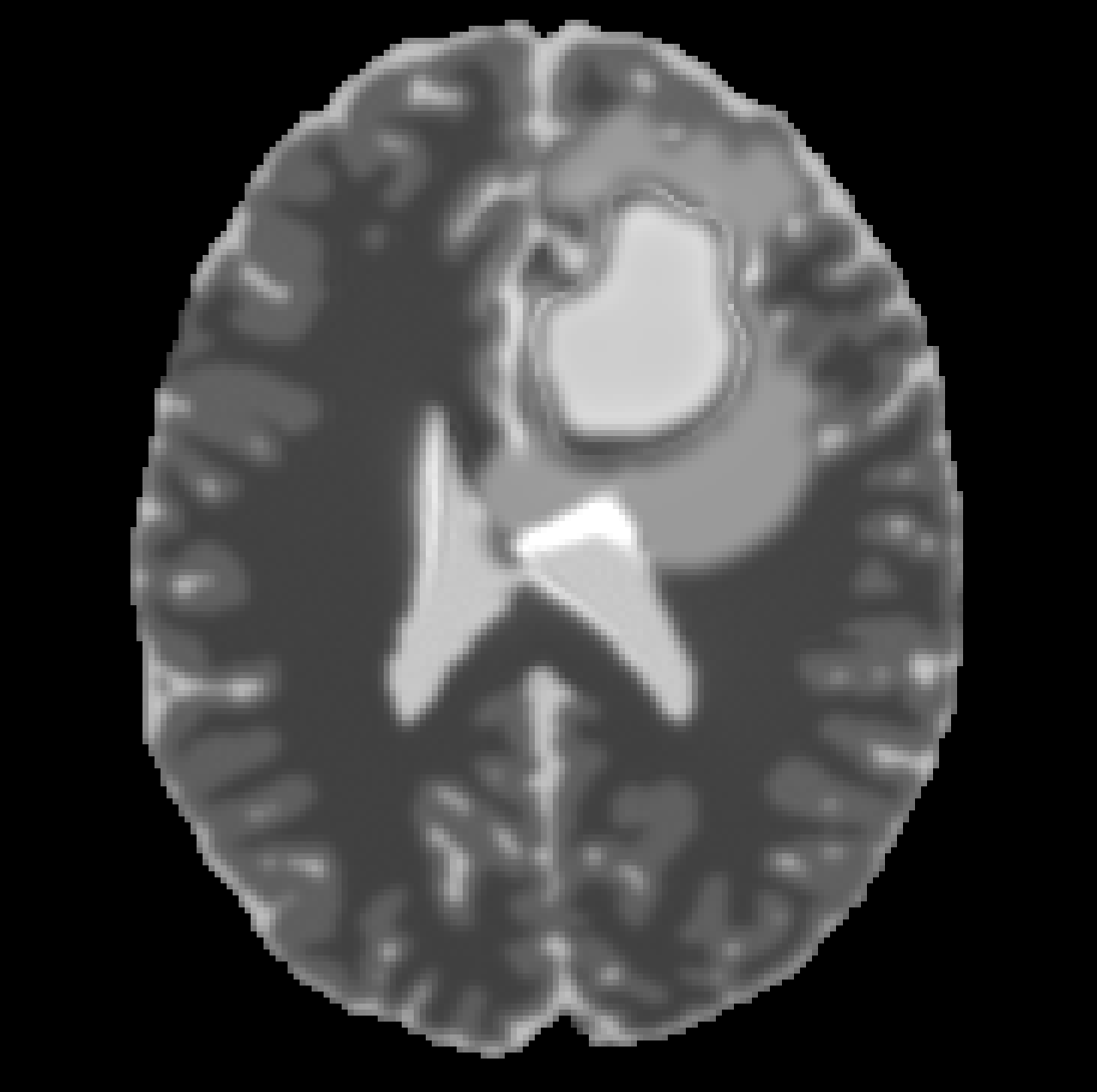}
		\caption{}
	\end{subfigure}%
	\begin{subfigure}{0.2\linewidth}
		\centering
		\includegraphics[height=1\linewidth, width=1\linewidth]{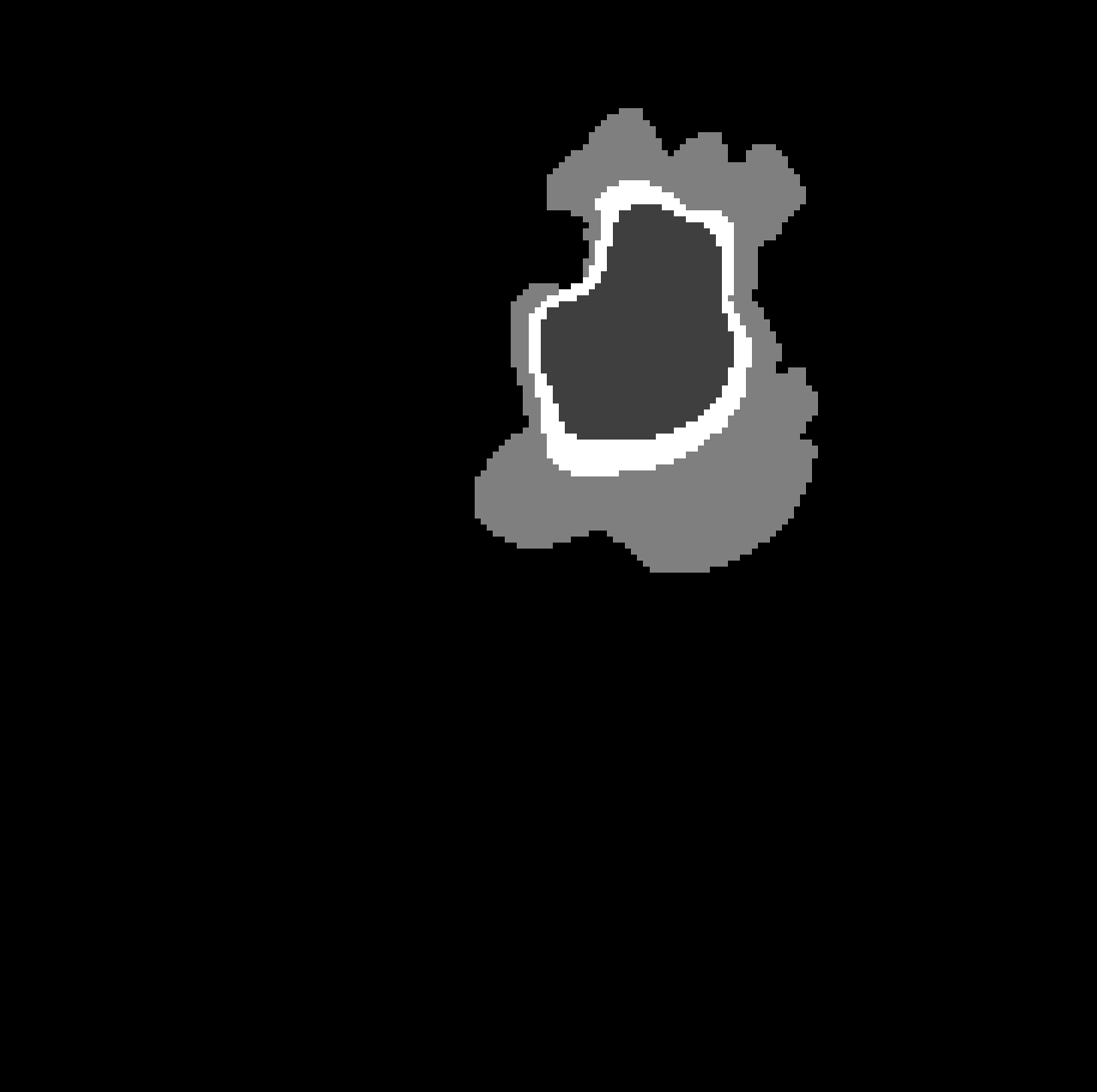}
		\caption{}
	\end{subfigure}
	\caption{Illustration of the phenomenological structures of a glioblastoma from MRI scans. All images are 2D slices of 3D volumetric datasets (\textit{Top row}) These images correspond to a specific axial slice of \textit{(a)} FLAIR, \textit{(b)} T1, \textit{(c)} T1-Gd, \textit{(d)} T2 MRI and \textit{(e)} segmentation of a real patient brain, taken from the Multimodal Brain Tumor 
		Segmentation Challenge, 2017~\citep{menze2015multimodal,bakas-davatzikos-e17b} training dataset. The peritumoral edema (\textit{light gray segmentation}) is visible in the hyper-intense signal in the FLAIR image, while the enhancing tumor structures (\textit{white segmentation}) and necrotic tumor core (\textit{dark gray segmentation}) are visible in the T1-Gd MRI scan. We can also observe a noticeable amount of mass effect. (\textit{Bottom row}) Simulated MRI modalities from our current multispecies model with mass effect. These are generated by growing a synthetic tumor in silico using a healthy brain image. Once we solve our model, we create images by using correlations between tissue type and MRI intensity obtained by existing segmentations. We use the scans (and their corresponding segmentations) from the GLISTR dataset~\citep{gooya-biros-davatzikos-e12} to achieve this.
		As one can see, the simulation captures many of the important features present in a real MR image. 
	}.
	\label{f:intro_illustration}
\end{figure}

A single species reaction-diffusion PDE has been one of the most popular modeling frameworks~\citep{clatz2005realistic,
	hogea2007modeling, hogea2008brain, jbabdi2005simulation,
	konukoglu2010image, mang2012biophysical, swanson2000quantitative,
	swanson2002virtual, rekik2013tumor}. This simple model attempts to capture two distinct
behaviors of malignant tumor growth: proliferation (reaction) and infiltration
(diffusion). However it captures neither the mass effect nor the distinct imaging characteristics of the visible tumor in MRI.
One of the most distinct features of glioblastomas is the presence of a proliferative rim of tumor cells surrounding a central necrotic core (dead tumor cells). The proliferative rim contains enhancing and non-enhancing tumor cells visible from different MRI contrasts. Moreover, imaging also highlights regions of peritumoral edema. Edema typically surrounds the proliferative rim and is known to be dispersed with highly migratory tumor cells, called infiltrative or invasive tumor cells. These cells, themselves, are not visible in MRI scans. They invade healthy parenchyma to distances that measure several centimeters beyond the detectable tumor core~\citep{giese2003}.
These cells are also able to invade even in the presence of treatment and can escape surgical resections, leading to recurrence~\citep{giese2003}. Thus, the extent of tumor invasion along with the sub-structures of a glioblastoma are highly important factors when planning treatment options and estimating survival times for the patient. An illustration of these imaging characteristics is shown in~\fref{f:intro_illustration}.

Our hypothesis is that by designing a  model that can capture these imaging characteristics (edema, enhancing and necrotic tumor, and mass effect), we will be able to extract clinically useful information. The model parameters can be inferred in a patient-specific manner and help improve the mathematical characterization of tumor, which ultimately  betters the clinical outcome. In this paper, our goal is modest. We introduce a possible model that integrates the structure of a glioblastoma with its mechanical effects on surrounding brain tissue. Multispecies models are useful in this context as they help in delineating the different tumor regions effectively without ad-hoc thresholding operations, which one might have to use when working with single species models. However, we also want to emphasize that we tried to design a model that it is as \emph{simple} as possible so it can be used in a robust way for parameter estimation. Indeed, highly complex, first principle models include many more tumor species, angiogenesis, chemotaxis, porous media that capture the interstitial fluid and extracellular matrix, and sophisticated models of growth. However, such models have a very large number of unknown parameters and pose outstanding numerical challenges with respect to both simulation and parameter estimation. Following the ideas in~\citep{yankeelov-miga13},  our goal here is to introduce a \emph{minimal} model that serves our clinical objective.

\textit{Related work:}  There have been varied approaches to modeling this phenomena on a molecular basis to a tissue (continuum) level~\citep{bellomo2008foundations, oden2013selection}. 

\begin{figure}
	\begin{subfigure}{0.25\linewidth}
		\centering
		\includegraphics[height=1.1\linewidth, width=1\linewidth]{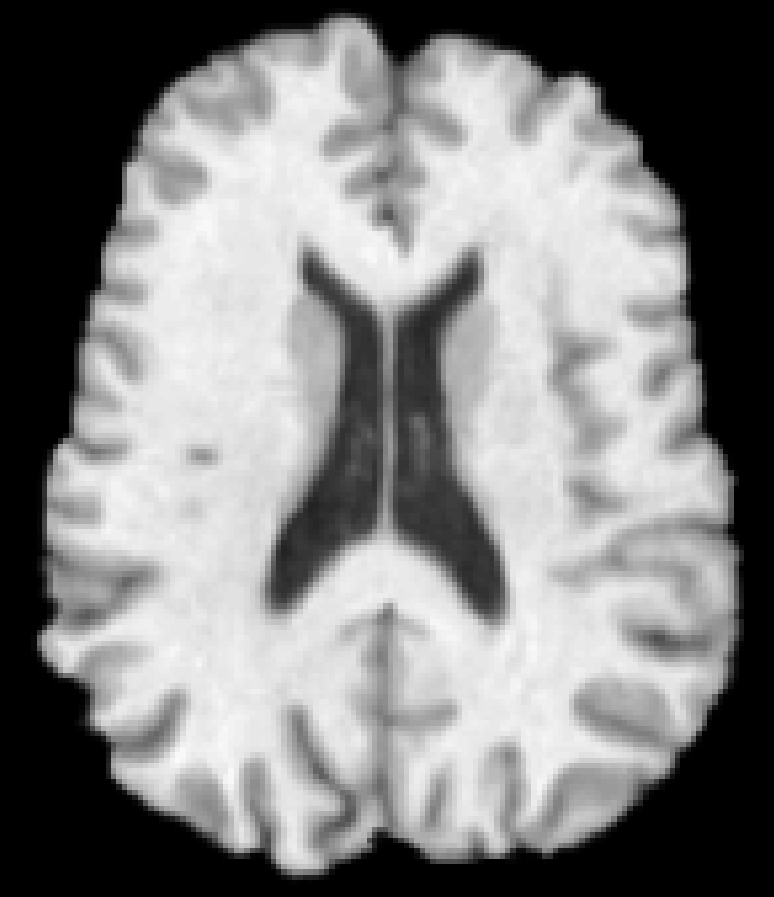}
		\caption{}
	\end{subfigure}%
	\begin{subfigure}{0.25\linewidth}
		\centering
		\includegraphics[height=1.1\linewidth, width=1\linewidth]{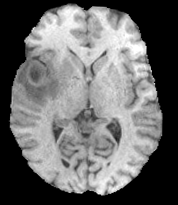}
		\caption{}
	\end{subfigure}%
	\begin{subfigure}{0.25\linewidth}
		\centering
		\includegraphics[height=1.1\linewidth, width=1\linewidth]{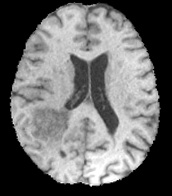}
		\caption{}
	\end{subfigure}%
	\begin{subfigure}{0.25\linewidth}
		\centering
		\includegraphics[height=1.1\linewidth, width=1\linewidth]{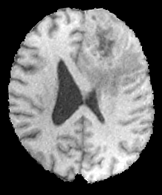}
		\caption{}
	\end{subfigure}
	\caption{T1 MRI scans of \textit{(a)} a healthy brain, \textit{(b)} brain with negligible mass effect, \textit{(c)} moderate mass effect and \textit{(d)} significant mass effect.}
	\label{f:me_real}
\end{figure}

The most widely used tumor-growth model is the single species, reaction-diffusion model~\citep{murray1989mathematical}. This model has proven to be quite effective in describing the whole tumor structure of glioblastomas~\citep{clatz2005realistic,	hogea2007modeling, jbabdi2005simulation,
	konukoglu2010image, mang2012biophysical, swanson2000quantitative,
	swanson2002virtual, rekik2013tumor}.
But tumors are very complex.  The underlying processes include mitosis, invasion, angiogenesis, biomechanics, environment quality, genotype, and gene expression. Approaches span from modeling each tumor phenotype on a cellular level~\citep{alarcon2003, gerlee2009, anderson2009} to macroscopic descriptions of tumor densities and nutrient supply \citep{HawkinsDaarud:2013a, konukoglu2010extrapolating, konukoglu2010image, swanson2008quantifying, Swanson:2011a,bellomo2008foundations, oden2013selection}).  One of the simplest multispecies models is based  on the “go-or-grow” hypothesis for differentiation. This hypothesis stems from experimental evidence~\citep{giese1996migration, giese2003} that suggests the existence of tumor cells in two states: one which proliferates quickly, but moves slowly (the proliferative one) and another which rapidly migrates but proliferates slowly (the invasive one). Further, tumor cells can mutate into the other phenotype based on the tumor microenvironment: cells become more migrating in a dearth of nutrients and proliferating in rich environments~\citep{hatzikirou2012}. While some models conform to the ``go-or-grow" hypothesis~\citep{saut2014multilayer, pham2012density}, others do not consider this phenotype. \cite{Swanson:2011a} stipulate the existence of normoxic and hypoxic tumor cells which migrate at the same rate. These are complex multispecies models  but they do not include mass effect, which is important for both low and high-grade glioblastomas. MRI scans of patient brains with different ranges of mass effect are shown in~\fref{f:me_real}.

\noindent Although tumor models date to the 1950s, models with mass effect are more recent. Early models \citep{mohamed2005finite, hogea06a} decoupled mass effect from tumor growth. The brain was modeled as an elastic material with external forces controlling the size of the tumor and displacements of surrounding tissue. More recent models, specifically the ones introduced in ~\cite{hogea2007modeling,rahman2017,hormuth2018}, couple tumor dynamics with elasticity equations. These models show flexibility in capturing complex/realistic tumor shapes and associated mass effect. These models, however, only deal with a single species of tumor cells. 

\noindent \textit{Contributions:} In this paper, we propose a new model for 
tumor growth dynamics of glioblastomas, a novel numerical scheme, and we 
present exemplary results. In particular:

\begin{enumerate}
	\item We propose a new go-or-grow, multispecies  model coupled with  
	an elasticity model for mass effect (\secref{s:fwd_model}). We model proliferative cells, invasive cells, necrotic cells and oxygen  concentration.  We use a new, mass-conserving  formulation that excludes the cerebrospinal fluid (which is not mass-conserved in the MR-defined control volume we use). In addition, we introduce a \emph{screened} elasticity model that  can better localize mass effect.
	
	\item We propose and test novel numerical schemes to discretize and solve the resulting model PDEs (\secref{s:num_methods}). Introducing a two-way coupling between the tumor evolution equations and linear elasticity results in challenging numerical problems as it leads to time dependence of various material and tumor properties. This is because the brain geometry changes as the tumor grows (due to the mass effect). Further, solving the linear elasticity equations is computationally challenging. The elasticity operator contains time and space varying coefficients, and the non-linearity of the tumor growth model adds to the numerical challenges. The numerical schemes used in our solver include stable pseudo-spectral methods, a semi-Lagrangian scheme for transport equations and preconditioners for the variable elasticity and diffusion equations.   
	
	\item In the results section~\secref{s:results}, we compare our model with a single species model with and without mass effect. We perform sensitivity analysis of our model parameters and finally present synthetic MR images highlighting the characteristic features of a glioblastoma.
\end{enumerate} 

Our model is inspired by~\cite{Hogea:2007b} for the elasticity model and~\cite{saut2014multilayer} for the multispecies model. We detail the differences of our model with these two models in Section \ref{s:fwd_model}.

\noindent \textit{Limitations:}
Phenomenological models can account for a wide range of complex phenomena. An important phenomenon is angiogenesis, which can be measured with perfusion MRI data. We do not include angiogenesis as we wanted to minimize the number of unknown parameters to the extent possible. However, even without angiogenesis
our model has many parameters, many of which are patient specific. This is related to the second limitation, which is lack of validation. One way to validate our model is to infer
model parameters from actual patients and examine the mismatch between our model predictions and real patient data (see~\cite{Gholami:2016a} for
inverse tumor problem formulations with single species reaction-diffusion models). The model we present here is much more complicated, and the latter approach is the focus of our immediate future work.

\section{Forward Tumor Mathematical Model}
\label{s:fwd_model}
In the following subsections, we first introduce the screened elasticity model for mass effect, second we introduce the mass-conserving single species model with elasticity coupling and finally, the overall multispecies model. We also use the single species model for comparisons with our multispecies model.

\subsection{Elasticity model} \label{s:elasticity}
We model the displacement ($\vu$) due to tumor-induced mass effect using linear elasticity. The governing equations of
linear elasticity for an isotropic medium are given by
\begin{subequations}
	\begin{gather}
	\vect{T} = \lambda tr(\vE)\vect{I} +  2\mu\vE, \label{e:hook} \\
	\nabla \cdot (\mu (\igrad\vu +  \igrad\vu^T)) + \igrad(\lambda\nabla \cdot\vu) + \vect{b} = \vzero \quad \text{in}\;\;\Omega,
	\label{e:variable_elasticity}
	\end{gather}
\end{subequations}
where $\vect{T}$ is the stress tensor, $\vect{E}$ is the infinitesimal strain tensor, $(\lambda, \mu)$ are Lam\'e coefficients and $\vect{b}$ is the total body force.  In order to prevent excessive shear to limit far field effects of the resulting force, our model supports the screening of the elasticity equation with a screening coefficient, $\eta(\vx, t)$, which will be a 
function of tumor concentration. We take $\eta$ to be zero in the presence of
tumor cells, and a high value elsewhere to screen the effects of the
tumor. We show examples of the effects of screening in~\secref{s:results}.  We write the screened elasticity equation in a more compact form as:
\begin{equation}\label{e:screening}
-\eta \vu + \mathcal{L}\vu = \vect{b} \quad \mbox{in $\Omega$ },
\end{equation}
where by $\mathcal{L}$ we denote the linear elasticity operator.
The screening coefficient $\eta$ can be varied to obtain different mass effect ranges and subsequent  deformations of the material properties.
The Lam\'e coefficients $(\lambda(\vx, t),\mu(\vx, t))$ are computed as:
\begin{subequations}\label{e:lame}
	\begin{align}
	\mu &= \sum_s \frac{E_s}{2(1 + \nu_s)} \iota_s \\
	\lambda &= \sum_s \frac{\nu_sE_s}{(1 - 2\nu_s)(1 + \nu_s)} \iota_s.
	\end{align}
\end{subequations}
Here, $\iota_s (\vx, t)$ is the concentration of the constituent species $s$ (tumor and healthy cells), $E_{
	s}$ is the Young's modulus of the species $s$ and $\nu_{s}$ is Poisson's ratio of the species $s$.

\subsection{Reaction-diffusion model coupled with linear elasticity}
The single species tumor model consists of a conservation equation with reaction and diffusion source terms for the evolution of the tumor cell concentration, $c(\vx,t)$, and is coupled with the linear elasticity equation (Eq. \ref{e:screening}). The model can be summarized by the following equations: 
\begin{subequations}\label{e:rd}
	\begin{align}
	\partial_t c + \nabla \cdot (c\vu_t)- \mathcal{D}c - \mathcal{R}  &= 0 \quad \mbox{in $\Omega\times(0, 1]$ } \label{e:reac_diff}\\
	c_0 - \Phi \vp &= 0 \quad \mbox{in $\Omega$ }\\
	-\eta \vu + \mathcal{L}\vu - \vect{b} &= 0\quad \mbox{in $\Omega\times(0, 1]$ } \label{e:linear_elasticity} \\
	\partial_t g + \nabla \cdot (g\vu_t) + \frac{g}{g+w} \left( \mathcal{D}c + \mathcal{R}\right) &= 0 \quad \mbox{in $\Omega\times(0, 1]$ } \label{e:gm}\\
	\partial_t w + \nabla \cdot (w\vu_t)  + \frac{w}{g+w} \left( \mathcal{D}c + \mathcal{R}\right) &= 0 \quad \mbox{in $\Omega\times(0, 1]$ } \label{e:wm}\\
	\partial_t f + \nabla f \cdot \vu_t  &= 0 \quad \mbox{in $\Omega\times(0, 1]$ } \label{e:csf}\\
	g - g_0 &= 0 \quad \mbox{in $\Omega$} \\
	w - w_0 &= 0 \quad \mbox{in $\Omega$} \\
	f - f_0 &= 0 \quad \mbox{in $\Omega$}.
	\end{align}
\end{subequations}

\begin{center}
	\def\arraystretch{1.2}
	\begin{table}[!htbp]
		\caption{Common notations for the reaction-diffusion tumor model.}
		\label{t:rd_notations}
		\centering
		\begin{tabular}{ m{4cm}|m{7cm}} 
			\hline
			\textbf{Notation} & \textbf{Description} \\  
			\hline
			$c(\vx, t)$								&			Tumor concentration \\
			\ghligh
			$g(\vx, t)$								&			Gray matter concentration \\
			$w(\vx, t)$								&		   White matter concentration \\
			\ghligh
			$f(\vx, t)$								&			 Cerebrospinal fluid density \\
			$(c_0(\vx), g_0(\vx), w_0(\vx), f_0(\vx))$	&		   Initial conditions for corresponding cell densities \\
			\ghligh
			$\vu_t(\vx, t)$						& 			  Advection velocity \\
			$\vect{b}(\vx, t)$							&		   Linear elasticity forcing function \\
			\ghligh
			$\mathcal{D}$ (see ~\eref{e:diffusion_op})			 &				Diffusion operator \\
			$\mathcal{R}$ (see~\eref{e:reac_op})				 &				Reaction operator \\			
			\ghligh
			\hline
			$\vx = (x,y,z)$			&				Spatial location \\
			$t$							&				 Time \\
			\ghligh
			$\Omega$ 				&			   Spatial domain \\
			$\Phi$						&			  Gaussian basis functions for tumor initial condition parameterization  \\
			\ghligh
			$\vp\in \R^{N_p}$		&             Tumor inital condition parameterization \\
			\hline
		\end{tabular}		
	\end{table}
\end{center}
The basic model notation is described in \tref{t:rd_notations}. We briefly explain the individual components of the model below.

Our goal is to be able to decide the properties of a voxel in order to create a segmentation that we can compare with MRI imaging data. Therefore, we introduce the following assumption: the total cell density is mass-conserved (similar to the assumptions introduced in~\cite{saut2014multilayer}). We define the total cell density as a sum of all component densities, i.e:
\begin{equation}
c + g + w = m,
\label{e:pw_mass_consv}
\end{equation}
where $m$ is the total cell density, which we assume is conserved. It thus satisfies:
\begin{equation}
\partial_t m + \nabla \cdot (m\vu_t)  = 0 \quad \mbox{in $\Omega\times(0, 1]$}. \label{e:global_mass}
\end{equation}
Then the conservation laws for the healthy cells, \eref{e:gm} and
\eref{e:wm}, follow from \eref{e:global_mass}. Notice that the cerebrospinal fluid is not
included. We apply a pure advection equation (\eref{e:csf}) to the
cerebrospinal fluid to account for possible leakage. This approach
differs from the one in~\cite{Hogea:2007b}, where the evolution of
material properties is not in conservative form and does not account
for the loss of healthy cells through tumor growth and invasion.

The differential operator, $\mathcal{D}$, is the inhomogeneous isotropic diffusion operator:
\begin{equation}\label{e:diffusion_op}
\mathcal{D}c = \div (\vK(g(\vx, t), w(\vx, t), f(\vx, t), \vx)\igrad c).
\end{equation}
The diffusion coefficient, $\vK$ is given by
\begin{equation}\label{e:diffusion_tensor}
\vK(g(\vx, t),w(\vx, t), \vx) = k_g g(\vx, t) \vI + k_w w(\vx, t) \vI,
\end{equation}
where $k_g$ and $k_w$ are the constant diffusion rates in gray and white
matter, respectively.  
The reaction operator is a non-linear logistic growth function: 
\begin{equation}\label{e:reac_op}
\mathcal{R} = \rho(\vx, t) c(1 - c),
\end{equation}
where
\begin{equation}\label{e:reaction_op}
\rho(\vx, t) = \rho_g g(\vx, t) + \rho_w w(\vx, t).
\end{equation}
Here, $\rho_g$ and $\rho_w$ are constant growth rates in gray and white matter, respectively. Note that the
diffusion and reaction operators now depend on the material properties that
change in time. This is due to the evolution of the tumor growth, which in turn
displaces the surrounding tissue.

The forcing function for the linear elasticity equations is modeled as
\begin{equation}\label{e:force}
\vect{b} =\zeta \text{tanh}(c)\igrad c,
\end{equation}
where $\zeta$ is a constant.  The choice of the force function is not unique and other formulations are possible as well~\citep{Hogea:2008b}.  Here, we assume that the force exerted on the brain tissue is proportional to tumor concentration gradient. The addition of the $\text{tanh}(c)$ term is to enforce small displacement forces where the tumor concentration is small. More complex models like poroelasticity or growth models~\citep{goriely2011}, which change the constitutive equation and write the deformation gradient as a product of an elastic and growth term  can be used, but these are highly nonlinear and with a large number of unknown parameters. Since here we are not developing a first-principles model but instead a more phenomenological model to be used in conjunction with imaging information, we use the simple linear elasticity model of~\secref{s:elasticity}. 

Regarding boundary conditions, we assume zero tumor flux on the skull and cerebrospinal fluid boundaries and zero displacement on the skull.
The Lam\'e coefficients
are different depending on the tissue type (tumor, gray matter, white
matter, cerebrospinal fluid). In our model (both single and
multi-species), we assume that the healthy tissues of the brain are
slightly compressible and the tumorous tissues are nearly
incompressible with a Young's modulus similar to that of healthy
tissues.  The cerebrospinal fluid is modeled as a highly compressible
and soft material. The different model parameters used for our
simulations are highlighted in \tref{t:setup}.

\subsection{Multispecies model coupled with linear elasticity}
We modify the model introduced by~\cite{saut2014multilayer} and couple it with
linear elasticity equations to capture mass effect. The basic structure of~\cite{saut2014multilayer} assumes that active tumor cells to exist in either one of two states, proliferative and invasive. If the tumor microenvironment has sufficient concentration of oxygen and other nutrients, the tumor cells grow through rapid mitosis by consuming those nutrients. If the oxygen concentration becomes low (hypoxia), the cells switch from proliferative to invasive. Invasive cells migrate to surrounding areas with higher nutrition concentration and switch back to proliferating cells when such an environment becomes available. This model assumes the only significant environmental factor affecting the state of tumor cells to be oxygen. In the event of severe hypoxia, the tumor cells die and become necrotic, typically located in the center of the tumor. The model is given by the following set of partial differential equations. 

\begin{center}
	\def\arraystretch{1.2}
	\begin{table}[!htbp]
		\caption{Common notations for the multispecies go-or-grow tumor model.}
		\label{t:gog_notations}
		\centering
		\begin{tabular}{ m{3cm}|m{8cm}} 
			\hline
			\textbf{Notation} & \textbf{Description} \\  
			\hline
			$p(\vx,t)$						&		    Proliferative tumor cell concentration \\
			\ghligh
			$i(\vx,t)$						&			 Invasive tumor cell concentration \\
			$n(\vx,t)$						&		   Necrotic tumor cell concentration \\
			\ghligh
			$o(\vx,t)$					&			Oxygen concentration \\
			$o_\text{heal}$  &				Healthy cell oxygen concentration \\
			\ghligh
			$\alpha(\vx,t)$ (see Eq. \eqref{e:alpha})&	transition rate from $p$ to $i$ cells \\
			$\beta(\vx,t)$ (see Eq. \eqref{e:beta}) & transition rate from $i$ to $p$ cells \\
			\ghligh
			$h(\vx,t)$ (see Eq. \eqref{e:death})	&			Oxygen threshold function \\			 
			$\mathcal{R}$ (see Eq. \eqref{e:gog_reac_op})	&			Proliferative cell reaction operator \\
			\ghligh
			$\mathcal{\tilde{R}}$	(see Eq. \eqref{e:gog_reac_inv_op})				&			Invasive cell reaction operator \\
			$\delta_p$	&		 Oxygen consumption rate \\
			\ghligh
			$\delta_s$ &		Oxygen supply rate \\			
			\hline
		\end{tabular}		
	\end{table}
\end{center}
\begin{subequations}\label{e:gog}
	\begin{align}
	\partial_t p + \div (p\vu_t) - \mathcal{R} + \alpha p - \beta i + h p  &= 0 \quad \mbox{in $\Omega\times(0, 1]$ }  \label{e:p}\\
	p_0 - \Phi \vp &= 0 \quad \mbox{in $\Omega$ }\\ 
	\partial_t i + \div (i\vu_t) - \mathcal{D}i -\mathcal{\tilde{R}}+\beta i - \alpha p +  h i &= 0 \quad \mbox{in $\Omega\times(0, 1]$ } \label{e:i}\\
	i_0 - \Phi \vect{\tilde{p}} &= 0 \quad \mbox{in $\Omega$ }\\
	\partial_t n + \div (n\vu_t) -hp -hi -h(g + w) &= 0 \quad \mbox{in $\Omega\times(0, 1]$ } \label{e:n}\\
	n_0  &= 0 \quad \mbox{in $\Omega$ } \\
	\partial_t o - \mathcal{D}o + \delta_pp - \delta_s(o_\text{heal} - o)(g + w) &= 0 \quad \mbox{in $\Omega\times(0, 1]$ } \label{e:ox}\\
	o_0 - o_\text{heal} &= 0 \quad \mbox{in $\Omega$ } \label{e:o}\\
	-\eta \vu + \mathcal{L}\vu - \vect{b} &= 0 \quad \mbox{in $\Omega\times(0, 1]$ } \label{e:gog_elasticity}\\
	\partial_t g + \nabla \cdot (g\vu_t) + \frac{g}{g+w} \left( \mathcal{D} + \mathcal{R} +\mathcal{\tilde{R}}\right) +hg &= 0 \quad \mbox{in $\Omega\times(0, 1]$ } \label{e:gm_wm}\\
	\partial_t w + \nabla \cdot (w\vu_t)  + \frac{w}{g+w} \left( \mathcal{D} + \mathcal{R} +\mathcal{\tilde{R}}\right) +hw &= 0 \quad \mbox{in $\Omega\times(0, 1]$ } \label{e:w}\\
	\partial_t f + \nabla f \cdot \vu_t  &= 0 \quad \mbox{in $\Omega\times(0, 1]$ } \\
	g - g_0 &= 0 \quad \mbox{in $\Omega$} \\
	w - w_0 &= 0 \quad \mbox{in $\Omega$} \\
	f - f_0 &= 0 \quad \mbox{in $\Omega$}.
	\end{align}
\end{subequations}

The common notations used in the multispecies model are outlined in \tref{t:gog_notations}. We provide a brief description of the details of the model below.

The governing equation for proliferative cells is a conservation equation with the following source terms: reaction corresponding to cell mitosis in favorable environments, phenotype switches between proliferative and invasive based on the quality of the environment and a death term in hypoxic regions. The evolution of invasive cells is governed by a diffusion equation with source terms representing similar behavior as proliferative cells. The conservation equation for necrotic cells is primarily driven by sources corresponding to the death of tumorous and healthy cells in hypoxic environments. 

Ignoring cerebrospinal fluid, we define the total cell density as:
\begin{equation}
p + i + n + g + w = m.
\label{e:pw_mass_consv_gog}
\end{equation}

The conservation laws for the healthy cells follow from the mass conservation of the total cell density (similar to the  single species model). We use a pure advection equation to model the evolution of the cerebrospinal fluid. 

We use a thresholding function $h$ based on the concentration of oxygen to model the death of cancer and healthy cells through a death rate $\gamma$: 

\begin{equation}\label{e:death}
h = \gamma \mathcal{H}(o_{\text{hypoxia}} - o),
\end{equation}

where $o_{\text{hypoxia}}$ is the hypoxia threshold and $\mathcal{H}$ is a smoothed Heaviside function. We model the reaction operator for the proliferative cells $\mathcal{R}$ as:

\begin{eqnarray}\label{e:gog_reac_op}
\mathcal{R} = 
\begin{dcases}
\rho p(1 - p),& o > o_{\text{inv}}\\
\left( \frac{o - o_{\text{mit}}}{o_{\text{inv}} - o_{\text{mit}}} \right) \rho p(1- p),&  o_{\text{inv}} \geq o \geq o_{\text{mit}}\\
0,& o <  o_{\text{mit}}, \\
\end{dcases}
\end{eqnarray}

where $o_{\text{mit}}$ and $o_{\text{inv}}$ are mitosis and invasive
oxygen thresholds, respectively. We use $o_{\text{mit}} = (o_{\text{hypoxia}} + o_{\text{inv}}) / 2$. Here, $\rho = \rho(\vx, t)$ is the proliferation rate as defined in~\eref{e:reac_op}.
For invasive cells, the reaction operator $\mathcal{\tilde{R}}$ is similarly defined as
\begin{eqnarray}\label{e:gog_reac_inv_op}
\mathcal{\tilde{R}} = 
\begin{dcases}
\rho i(1 - i),& o > o_{\text{inv}}\\
\left( \frac{o - o_{\text{mit}}}{o_{\text{inv}} - o_{\text{mit}}} \right) \rho i(1- i),&  o_{\text{inv}} \geq o \geq o_{\text{mit}}\\
0,& o <  o_{\text{mit}}. \\
\end{dcases}
\end{eqnarray}
We assume that invasive cells proliferate at a much smaller rate compared to proliferative cells. We take $\rho_w^i$, the proliferation rate of invasive cells in white matter as a small fraction of $\rho_w$ (see~\eref{e:reac_op}). 
We model the transition rate $\alpha$ as a decreasing function of oxygen
concentration by the expression:

\begin{equation}\label{e:alpha}
\alpha = \alpha_0 \mathcal{H}(o_{\text{inv}} - o). 
\end{equation}

The transition rate $\beta$ is modeled as: 

\begin{equation}\label{e:beta}
\beta  = \beta_0(\mathcal{H}(\sigma_{b} - p - i)o),
\end{equation}

where $\sigma_{b}$ is a threshold above
which the transition to $p$ is prohibited. The rate is an increasing function 
of oxygen concentration to prevent cells from converting to the proliferative 
phenotype in scarcity of oxygen. 

The evolution of oxygen is modeled through a diffusion equation. The source
of oxygen is assumed to be proportional to the concentration of healthy cells and oxygen is 
consumed by proliferation at a constant rate $\delta_p$. 

We use
\begin{equation}\label{e:forcingfunction}
\vect{b} =\zeta\text{tanh}(p + n)\igrad (p + n).
\end{equation}
for the forcing function of the linear elasticity, similar to the single-species model. 

\subsection{Tumor-associated brain edema model}

Cerebral edema in glioblastomas arise primarily from the leakage of protein and fluid
into the extra-cellular matrix of the brain. Edema is a prominent image phenotype of glioblastomas, since it is clearly visible in MR images and is typically infiltrated by invasive tumor cells that lead to post-resection recurrence. The mechanism of fluid accumulation due to the tumor is often modeled as a consequence of the infiltrative property of tumor cells. We choose a model based on the works of~\cite{HawkinsDaarud:2013a}. The infiltrative tumor cells cause the fluid to leak into the extra-cellular space, where it moves through diffusion. A constant drainage term models the re-absorption  of fluid into the vascular system. The equations governing the evolution of edema are one-way coupled to the multispecies tumor growth model and are given below. 
\begin{equation}\label{e:edema}
l_t = \mathcal{D} l + \delta_e \frac{i}{i + \delta_{\text{half}}} (1 - i) - \delta_l l,       
\end{equation}  
where $l$ is the edematous fluid concentration, $\delta_e$ is a measure of the
transmission rate of edema into the extra-cellular space, 
$\delta_{\text{half}}$ is the concentration of invasive
cells at which $\delta_e$ reaches half of its maximum value and $\delta_l$ is
the rate of drainage of edema back into the system. We note that the
nature of edematous fluid from MRI images can also be approximated by:
\begin{equation}
l = (1 - p - i - n) \mathcal{H} (i - i_{\text{threshold}}).
\end{equation}
with some threshold value chosen for invasive cells, $i_{\text{threshold}}$.

\subsection{Model parameters}\label{e:parameters}

The multi-species model presented above has 27 parameters (essentially material properties).
Approximate (range of) values are give for some of these parameteres in the literature, and we use those values from the literature. 
The values for the various model parameters are summarized in \tref{t:setup}.
In particular, we refer to~\cite{Gholami:2016a} for reaction and diffusion coefficient values,~\cite{saut2014multilayer} for the multispecies model parameter values,~\cite{Hogea:2007b} for the elasticity model parameter values and~\cite{HawkinsDaarud:2013a} for the edema model parameter values. For screening coefficient values and edema model parameter values, we experiment with different values to produce reasonable qualitative characteristics of mass effect and edema in MRI images.

\begin{center}
	\def\arraystretch{1.2}
	\begin{table}[!htbp]
		\caption{Model parameters used in the forward simulations. For the screening parameters, we experiment with different values to obtain the best results (see ~\secref{s:sens} for other range of values that can be used and the sensitivity of the model to these values). For the edema model parameters, we refer to~\cite{HawkinsDaarud:2013a} and experiment with different values to produce the qualitative characteristics of edema in MRI images.}
		\label{t:setup}
		\centering
		\begin{tabular}{ m{9cm}|m{2cm}} 
			\hline
			\textbf{Parameter} & \textbf{Value} \\  
			\hline
			\hline
			\multicolumn{1}{l}{\textit{Reaction and diffusion coefficients} (see ~\cite{gholami-mang-biros15})}  \\  
			\hline
			\ghligh
			Diffusion coefficient in white matter, $k_w$  \hspace{4pt}(see Eq. \eqref{e:diffusion_tensor}) & 0.1 \\
			Diffusion coefficient in gray matter, $k_g$  \hspace{4pt}(see Eq. \eqref{e:diffusion_tensor}) & 0 \\
			\ghligh
			Reaction coefficient in white matter, $\rho_w$  \hspace{4pt}(see Eq. \eqref{e:reaction_op})  & 8 \\	
			Reaction coefficient in gray matter, $\rho_g$  \hspace{4pt}(see Eq. \eqref{e:reaction_op})  & 0 \\	
			\ghligh
			Reaction coefficient of invasive cells in white matter, $\rho_w^i$  \hspace{4pt}(see Eq. \eqref{e:gog_reac_inv_op})  & 0.8 \\	
			\hline
			\multicolumn{1}{l}{\textit{Go-or-grow model parameters} (see ~\cite{saut2014multilayer})}   \\  
			\hline		
			Transition rate coefficient from $p$ to $i$, $\alpha_0$ \hspace{4pt}(see Eq. \eqref{e:alpha})& 0.15 \\
			\ghligh
			Transition rate coefficient from $i$ to $p$, $\beta_0$ \hspace{4pt}(see Eq. \eqref{e:beta})& 0.02 \\		
			Death rate, $\gamma$ \hspace{4pt}(see Eq. \eqref{e:death})& 1 \\
			\ghligh
			Transition threshold from $i$ to $p$, $\sigma_{b}$\hspace{4pt}(see Eq. \eqref{e:beta}) & 0.9 \\
			Hypoxia oxygen threshold, $o_{\text{hypoxia}}$\hspace{4pt}(see Eq. \eqref{e:death}) & 0.65 \\
			\ghligh
			Invasive oxygen threshold, $o_{\text{inv}}$\hspace{4pt}(see Eq. \eqref{e:gog_reac_op}) & 0.7 \\
			Oxygen supply, $\delta_s$\hspace{4pt}(see Eq. \eqref{e:ox}) & 55 \\
			\ghligh
			Oxygen consumption rate, $\delta_p$\hspace{4pt}(see Eq. \eqref{e:ox})  & 8 \\
			
			\hline
			\multicolumn{1}{l}{\textit{Screening parameters}}    \\  
			\hline
			Screening factor in tumor cells, $\eta_{\text{tumor}}$\hspace{4pt}(see Eq. \eqref{e:screening})  & 0 \\
			\ghligh
			Screening factor in healthy cells, $\eta_{\text{healthy cells}}$\hspace{4pt}(see Eq. \eqref{e:screening}) & 10000 \\
			Screening factor in background, $\eta_{\text{background}}$ \hspace{4pt}(see Eq. \eqref{e:screening})& 10$^6$ \\	 
			\hline
			\multicolumn{1}{l}{\textit{Elasticity material properties} (see ~\cite{hogea2007modeling})}  \\  
			\hline
			\ghligh
			Young's modulus of gray matter, $E_{\text{gm}}$ (Pa)\hspace{4pt}(see Eq. \eqref{e:lame})	&	2100  \\
			Young's modulus of white matter, $E_{\text{wm}}$ (Pa)\hspace{4pt}(see Eq. \eqref{e:lame})	&	2100 \\
			\ghligh
			Young's modulus of cerebrospinal fluid, $E_{\text{csf}}$ (Pa)\hspace{4pt}(see Eq. \eqref{e:lame})	&	100  \\
			Young's modulus of tumor, $E_{\text{tumor}}$ (Pa)\hspace{4pt}(see Eq. \eqref{e:lame})	&	2100 \\
			\ghligh
			Young's modulus of background material, $E_{\text{bcg}}$ (Pa)\hspace{4pt}(see Eq. \eqref{e:lame})	&	15000  \\
			Poisson's ratio of gray matter, $\nu_{\text{gm}}$ \hspace{4pt}(see Eq. \eqref{e:lame})	&	0.3 \\
			\ghligh
			Poisson's ratio of white matter, $\nu_{\text{wm}}$ \hspace{4pt}(see Eq. \eqref{e:lame})	&	0.3 \\
			Poisson's ratio of cerebrospinal fluid, $\nu_{\text{csf}}$ \hspace{4pt}(see Eq. \eqref{e:lame})	&	0.1 \\
			\ghligh
			Poisson's ratio of background material, $\nu_{\text{bcg}}$\hspace{4pt}(see Eq. \eqref{e:lame}) 	&	0.48 \\
			Poisson's ratio of tumor, $\nu_{\text{tumor}}$\hspace{4pt}(see Eq. \eqref{e:lame}) 	&	0.45 \\
			\ghligh
			Forcing function constant, $\zeta$	\hspace{4pt}(see Eq. \eqref{e:force})			& 			 40000 \\
			\hline
			\multicolumn{1}{l}{\textit{Edema model parameters}}\\  
			\hline
			Transmission rate of edematous fluid into extra-cellular space, $\delta_e$ \hspace{4pt}(see Eq. \eqref{e:edema}))	&	40 \\
			\ghligh
			Half-max invasive concentration, $\delta_{\text{half}}$  \hspace{4pt}(see Eq. \eqref{e:edema}))	&	0.01 \\
			Drainage rate of edematous fluid, $\delta_l$  \hspace{4pt}(see Eq. \eqref{e:edema}))& 80 \\		   
			\hline
		\end{tabular}
	\end{table}
\end{center}

\section{Discretization and numerical scheme}
\label{s:num_methods}
We use the Strang operator splitting method~\citep{strang1968construction} in conjunction with pseudospectral methods to numerically solve the non-linear system of PDEs. First, we describe the discretization scheme for the single species model outlined in \eref{e:rd}. Then, we describe the discretization for the multispecies model.

We use a fictitious domain method where the brain is assumed to reside in a cubic box. The space is discretized uniformly into $256^3$ nodes with spatial resolution $1\text{mm}\times 1\text{mm}\times 1\text{mm}$.

Given tumor concentration $c^n$ at time step $n$, healthy tissue concentrations $(g^n, w^n, f^n)$, and corresponding material-dependent reaction and diffusion coefficients, we solve the single species model (\eref{e:rd}) through the following operator splitting steps:
\begin{itemize}
	\item Solve the advection equations $\partial_t q + \nabla \cdot (q \vu_t) = 0$ for $q = (c, g, w)$ over time $\Delta t$ using the semi-Lagrangian method (see~\cite{Mang:2016c, Falcone:1998a}),
	with $(c^n, g^n, w^n)$ as initial condition and current velocity $\vu_t^{n}$ to obtain $(c^{\dagger}, g^{\dagger}, w^{\dagger})$.
	\item Solve the advection equation $\partial_t f + \nabla f \cdot \vu_t  = 0$ for $f$ over time $\Delta t$ using the semi-Lagrangian method
	with $f^n$ as initial condition and current velocity $\vu_t^{n}$ to obtain $f^{n+1}$.
	\item Solve the diffusion equation $\partial_t c - \mathcal{D}c = 0$  over time $\Delta t$ using the Crank-Nicolson method (~\cite{crank1996})
	with $c^{\dagger}$ as initial condition to obtain $c^{\dagger \dagger}$.
	\item Solve the reaction equations $\partial_t q - \mathcal{R}_q = 0$ over time $\Delta t$ explicitly with $(c^{\dagger \dagger}, g^{\dagger}, w^{\dagger})$ as initial condition to obtain $(c^{n+1}, g^{n+1},w^{n+1})$. Here, $\mathcal{R}_q$ is the reaction/source operator for Eq. \ref{e:reac_diff}, Eq. \ref{e:gm} and  Eq. \ref{e:wm}. Update the reaction, diffusion and elasticity coefficients using the new healthy tissue and tumor concentration.
	\item Solve the variable linear elasticity equation (\eref{e:linear_elasticity}) using Krylov subspace methods with forcing function computed from  $c^{n+1}$ and Lam\'e coefficients computed from $(g^{n+1}, w^{n+1}, f^{n+1})$ to obtain $\vu^{n+1}$. Update velocity $\vu_t^{n+1}$ using backward time differencing.
\end{itemize}

Now, we provide specific details of each step. For the diffusion equation, we use a
pseudo-spectral spatial discretization with a Crank-Nicolson scheme in time.
The diffusion equation reduces to:

\begin{equation}
\left(  1 - \frac{\Delta t}{2}\mathcal{D} \right) c^{n+1} = \left(1 +  \frac{\Delta t}{2}\mathcal{D}\right)c^n.
\end{equation}

We solve this symmetric, implicit system of equations using the
Conjugate Gradient (CG) method.
We precondition this Krylov solver by solving the diffusion
equation using constant coefficients computed by averaging the inhomogeneous
diffusion coefficient over the spatial domain (similar to the works of ~\cite{gholami-mang-biros15}).  

All transport equations are solved using a semi-Lagrangian scheme, which is unconditionally stable for solving the linear advection equations. We use a second order in time and third order in space (for interpolation) to solve for the semi-Lagrangian trajectories of a scalar field, $\nu (\vx, t)$. Here, we briefly describe the semi- Lagrangian scheme using the following transport equation: 
\begin{equation}
\partial_t \nu + \nabla \nu \cdot \vu_t = g(\nu, \vx).
\end{equation}

In this method, we compute a new grid point $\vX$ using the scheme below:
\begin{subequations}
	\begin{align}
	\vX_* &= \vx - \Delta t \vu_t (\vx), \\
	\vX  &= \vx - \frac{\Delta t}{2} \left( \vu_t(\vx) + \vu_t(\vX_*)\right).
	\end{align}
\end{subequations}
We find the scalar field at the next time instant using:
\begin{subequations}
	\begin{align}
	\nu_*(\vx) &= \nu (\vX, 0) + \Delta t g(\nu(\vX, 0), X), \\
	g_*(\vx) &= g(\nu_*(\vx), \vx), \\
	\nu(\vx, \Delta t) &= \nu(\vX, 0) + \frac{\Delta t}{2} \left( g(\nu(\vX, 0), X) + g_*(\vx)\right).
	\end{align}
\end{subequations}
We use cubic interpolation to find field values at non-grid points
$\vX$ and $\vX_*$. Further details on the semi-Lagrangian method can be found
in~\cite{Mang:2016c}.

To compute the displacement $\vu$, we have to solve
\eref{e:linear_elasticity}. Since the Lam\'e coefficients are variable in the
domain, we use the Generalized Minimum Residual Method (GMRES) to iteratively
minimize the residual in the Krylov subspace of \eref{e:linear_elasticity}, up to a user defined
tolerance, $\tau$.  To increase the convergence rate, we precondition the variable elasticity equation
by solving it with constant Lam\'e coefficients computed by taking their
average over the domain. The constant elasticity coefficient equation can be
solved analytically as follows:

Using the identity of
$\div(\igrad\vu +\igrad\vu^T)= \Delta\vu + \igrad\div\vu$,
we need to solve the following equation:

\begin{equation}
\mathcal{L}\vu = \mu\ilap\vu + (\lambda+\mu)\igrad\div\vu = \vect{b}.
\end{equation}

Given the right hand side $\vect{b}$, we need to compute $\vu$.
Applying the Fourier transform on both sides, we obtain:

\begin{equation}
\left( \mu\vomega^T\vomega \vI + (\lambda+\mu)\vomega\vomega^T\right)\widehat{\vu} = \widehat{\vect{b}},
\end{equation}
where the hats denote the frequency domain, and $\vomega$ is the
corresponding wave numbers.  We, then, use
the Sherman-Morrison formula to compute $\vu$:
\begin{equation}
\vu = \mathcal{F}^{-1}\big((\frac{1}{\mu\vomega^T\vomega} - \frac{1}{(\mu\vomega^T\vomega)^2} \frac{(\lambda + \mu)\vomega\vomega^T}{1+\frac{\lambda+\mu}{\mu}})\widehat{\vect{b}}\big),
\end{equation}
where $\mathcal{F}^{-1}$ is the inverse Fourier transform. 
We enforce the zero tumor flux boundary condition and zero displacement boundary condition on the skull using appropriate smoothed penalized conditions using the penalty method (see~\cite{Hogea:2008b, del2003fictitious} for more details).

For qualitative assessments of mass effect, we compute principal stresses and maximum shear stress at every time step, using the stress tensor calculated from \eref{e:hook}. The principal stresses are visualized by the trace of the stress tensor and the maximum shear stress is computed from the Mohr's circle as:
\begin{equation}\label{e:princ_stress}
\tau_{\max} = \sqrt{\frac{(\vect{T}_{xx} - \vect{T}_{yy})}{2}^2 + \vect{T}_{xy}^2},
\end{equation}
where $T$ is given in~\eqref{e:variable_elasticity}. Further, for sensitivity of mass effect parameters like forcing factor $\zeta$, we compute the determinant of the deformation gradient (also known as the Jacobian $J$) as follows:
\begin{equation}\label{e:jacobian}
\mathcal{J} = \text{det}(I + \nabla u),
\end{equation}

For the multispecies model, we solve all the equations in a similar fashion as the single species model. We employ the following operator splitting steps:
\begin{itemize}
	\item Solve the advection equations $\partial_t q + \nabla \cdot (q \vu_t) = 0$ for $q = (p, i, n, g, w)$ over time $\Delta t$ using the semi-Lagrangian method
	with $q^n$ as initial condition and current velocity $\vu_t^{n}$ to obtain $(p^{\dagger}, i^{\dagger}, n^{\dagger}, g^{\dagger}, w^{\dagger})$.
	\item Solve the advection equation $\partial_t f + \nabla f \cdot \vu_t  = 0$ for $f$ over time $\Delta t$ using the semi-Lagrangian method
	with $f^n$ as initial condition and current velocity $\vu_t^{n}$ to obtain $f^{n+1}$.
	\item Solve the diffusion equations $\partial_t i - \mathcal{D}i = 0$ and $\partial_t o - \mathcal{D}o = 0$ over time $\Delta t$ using the Crank-Nicolson method
	with $i^{\dagger}$ and $o^n$ as initial conditions to obtain $i^{\dagger \dagger}$ and $o^{\dagger}$.
	\item Solve the reaction equations $\partial_t q - \mathcal{R}_q = 0$ over time $\Delta t$ explicitly with $(p^{\dagger}, i^{\dagger \dagger}, n^{\dagger}, g^{\dagger}, w^{\dagger})$ as initial condition to obtain $q^{n+1}$. Here, $\mathcal{R}_q$ is the reaction/source operator for Eq. \ref{e:p} -\ref{e:n} and  Eq. \ref{e:gm_wm} - \ref{e:w}. Update the reaction, diffusion and elasticity coefficients using the new healthy tissue and tumor concentration.
	\item Solve the reaction equation $\partial_t o + \mathcal{R}_o = 0$ over time $\Delta t$ explicitly with $o^{\dagger}$ as initial condition to obtain $o^{n+1}$.
	\item Solve the variable linear elasticity equation (\eref{e:gog_elasticity}) using Krylov subspace methods with forcing function and Lam\'e coefficients computed from $q^{n+1}$ to obtain $\vu^{n+1}$. Update velocity $\vu_t^{n+1}$ using backward time differencing.
\end{itemize}

The numerical parameters used are highlighted in \tref{t:numparams}. The typical number of Krylov solves for the diffusion equation is around one to three Conjugate Gradient iterations. For the preconditioned linear elasticity equations, the number of GMRES iterations is typically around 50 to 60. Preconditioning with the constant elasticity operator is quite effective and enables us to reduce the number of Krylov iterations from over 1000 to about 50. Also note that the spectral preconditioner is very cheap to apply since it only involves Fast Fourier transforms.

\begin{center}
	\def\arraystretch{1.2}
	\begin{table}[!htbp]
		\caption{Numerical parameters used in the forward simulations.}
		\label{t:numparams}
		\centering
		\begin{tabular}{ m{5cm}|m{4cm}} 
			\hline
			\textbf{Parameter} & \textbf{Value} \\  
			\hline
			Number of discretization points, $N_x=N_y =N_z$ 		& 			256 \\
			\ghligh
			Time step, $dt$													&			 0.01	 \\
			Tolerance for Krylov subspace solvers, $\tau$						&			0.001 \\
			\ghligh
			Time horizon										&				1 \\
			\hline
		\end{tabular}		
	\end{table}
\end{center}

\section{Numerical experiments}\label{s:results}
We perform a number of simulations to demonstrate qualitatively the behavior of our scheme, compare the single species with the multispecies model, illustrate the mass effect and the effects of screening, and explain how we can use our scheme to create synthetic MR images. In addition, we conduct a basic sensitivity analysis for a small number of parameters.

We perform all simulations using the BrainWeb
atlas~\citep{cocosco1997brainweb} (spatial resolution: $1mm\times 1mm\times 1mm$), which provides a realistic brain geometry segmented
into gray matter, white matter and cerebrospinal fluid. We use these values
as initial conditions for the evolution of material properties and
cerebrospinal fluid. To visualize the results of our simulations, we segment the brain into material properties
and tumor by choosing the label with maximum cell density at any voxel~\citep{gooya2011deformable}.
We overlay tumor concentrations onto these segmentations along with contours of the cerebrospinal
fluid at $t = 0$ to observe the evolution of tumor cells and tumor-induced deformations in material properties.

\begin{figure}[!htbp]
	\begin{subfigure}{.33\textwidth}
		\centering
		\includegraphics[height = 1\linewidth,width=1.\linewidth]{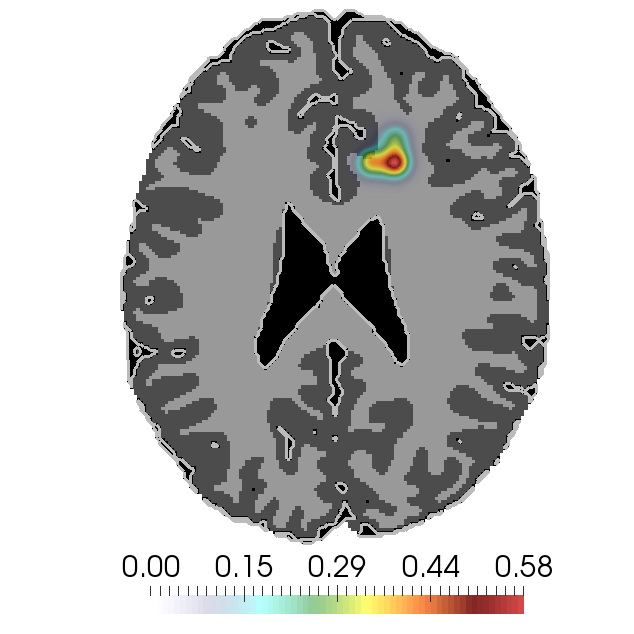}
		\caption{$t = 0$}
	\end{subfigure}%
	\begin{subfigure}{.33\textwidth}
		\centering
		\includegraphics[height = 1\linewidth,width=1.\linewidth]{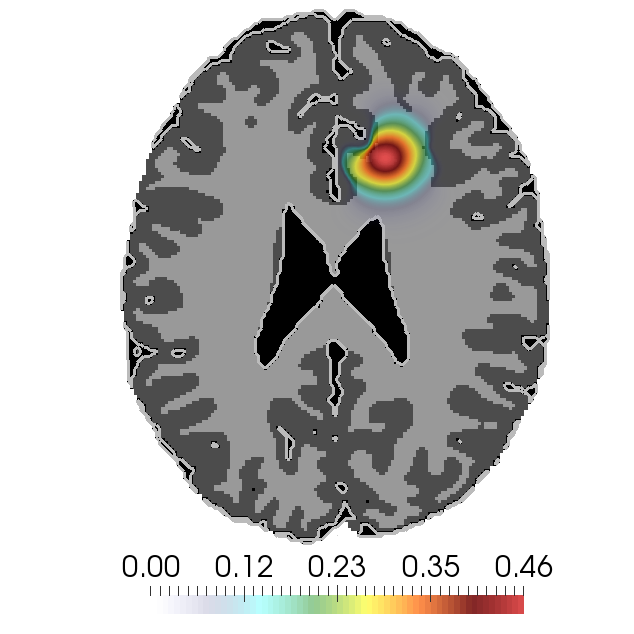}
		\caption{$t = 0.15$}
	\end{subfigure}%
	\begin{subfigure}{.33\textwidth}
		\centering
		\includegraphics[height = 1.\linewidth,width=1.\linewidth]{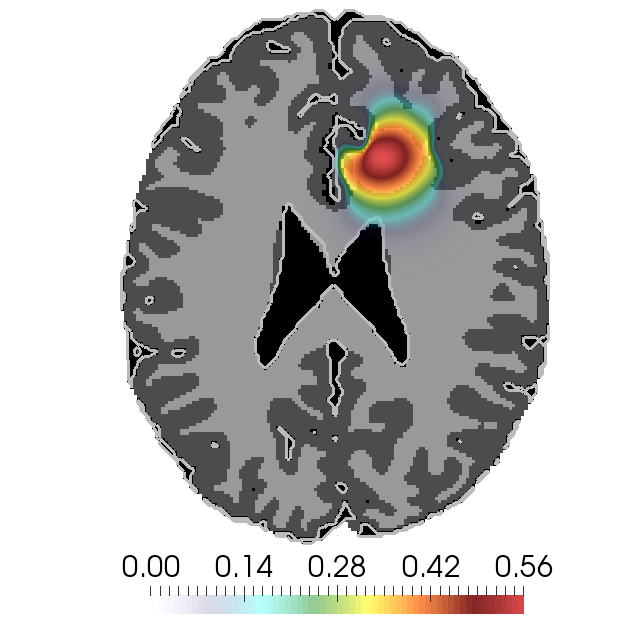}
		\caption{$t = 0.3$}
	\end{subfigure}
	\begin{subfigure}{.33\textwidth}
		\centering
		\includegraphics[height = 1\linewidth,width=1\linewidth]{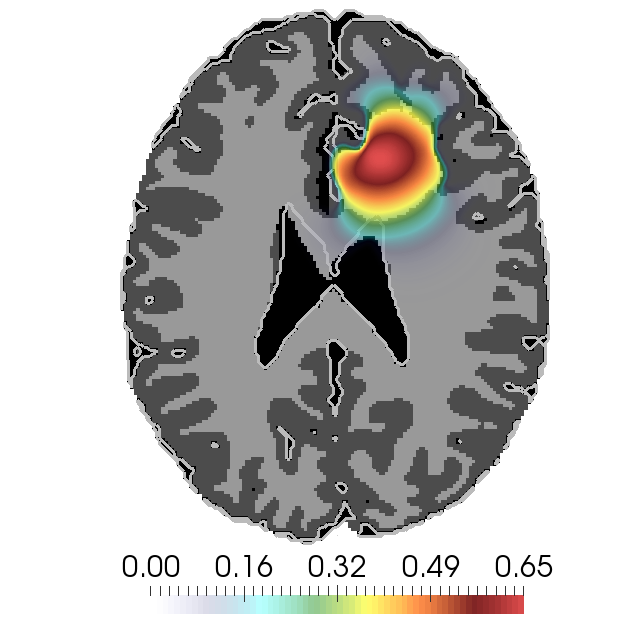}
		\caption{$t = 0.45$}
	\end{subfigure}%
	\begin{subfigure}{.33\textwidth}
		\centering
		\includegraphics[height = 1.\linewidth,width=1.\linewidth]{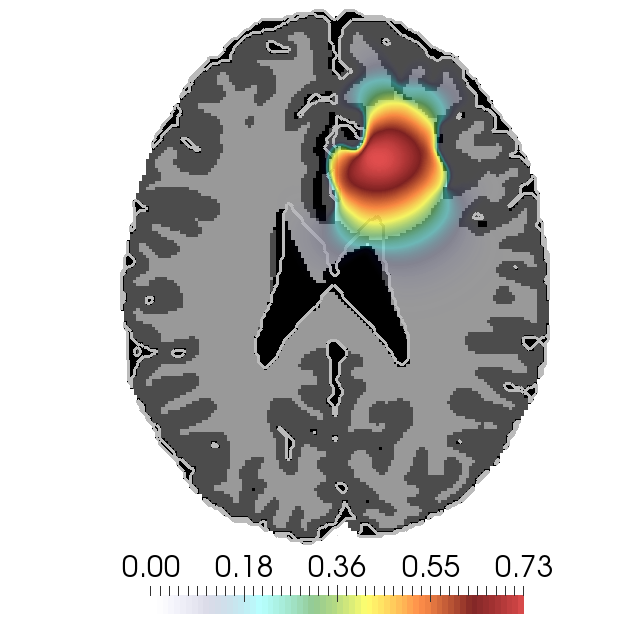}
		\caption{$t = 0.6$}
	\end{subfigure}%
	\begin{subfigure}{.33\textwidth}
		\centering
		\includegraphics[height = 1\linewidth,width=1.\linewidth]{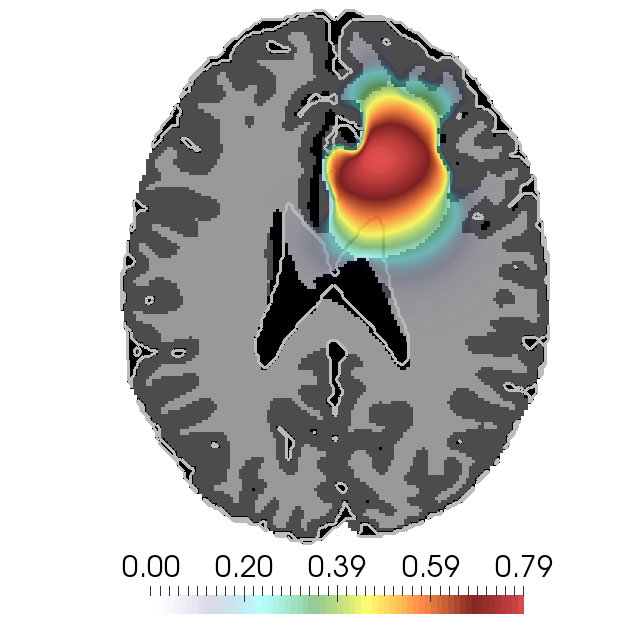}
		\caption{$t = 0.75$}
	\end{subfigure}
	\caption{Time evolution of segmentation of the brain and tumor concentration for the reaction-diffusion model with mass effect. The segmentations are overlayed with contours of the cerebrospinal fluid at initial time.
		As one can see, there is significant deformation of the cerebrospinal fluid and surrounding tissues.}
	\label{f:rd_seg}
\end{figure}

\begin{figure}[!htbp]
	\begin{subfigure}{.33\textwidth}
		\centering
		\includegraphics[height = 1.\linewidth,width=1.\linewidth]{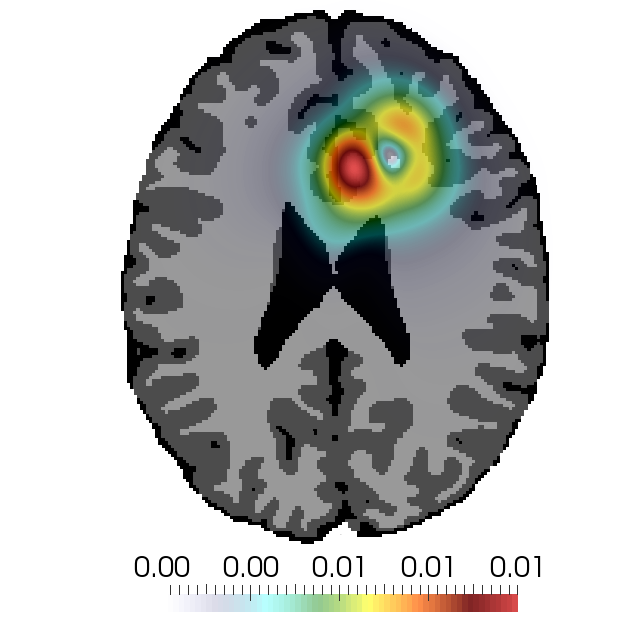}
		\caption{$t = 0$}
	\end{subfigure}%
	\begin{subfigure}{.33\textwidth}
		\centering
		\includegraphics[height = 1\linewidth,width=1.\linewidth]{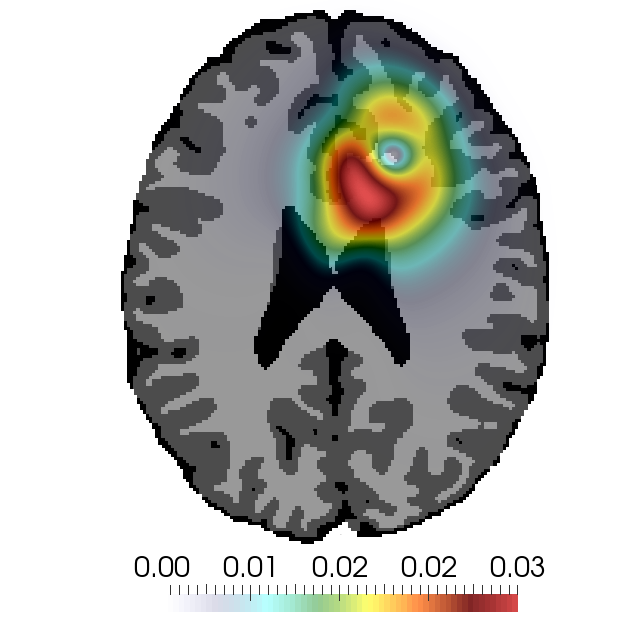}
		\caption{$t = 0.15$}
	\end{subfigure}%
	\begin{subfigure}{.33\textwidth}
		\centering
		\includegraphics[height = 1\linewidth,width=1.\linewidth]{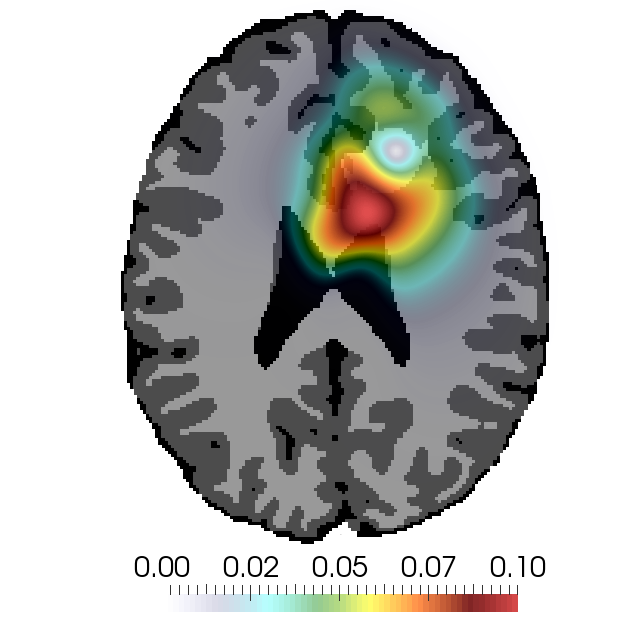}
		\caption{$t = 0.3$}
	\end{subfigure}
	\begin{subfigure}{.33\textwidth}
		\centering
		\includegraphics[height = 1.\linewidth,width=1\linewidth]{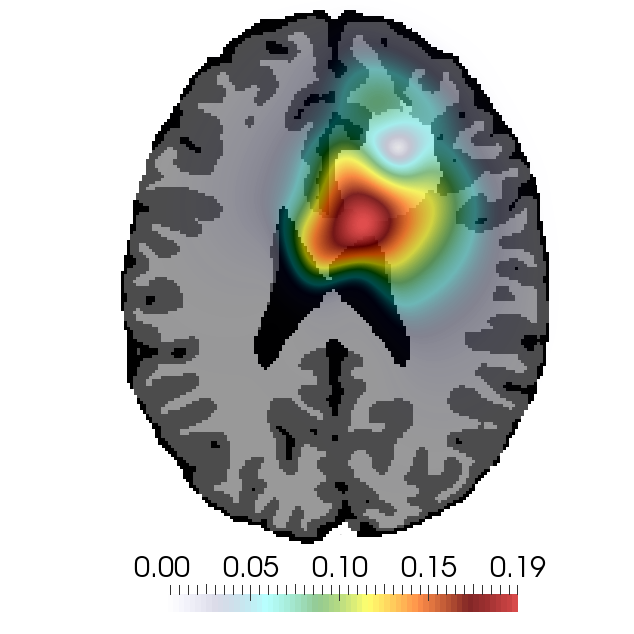}
		\caption{$t = 0.45$}
	\end{subfigure}%
	\begin{subfigure}{.33\textwidth}
		\centering
		\includegraphics[height = 1.\linewidth,width=1.\linewidth]{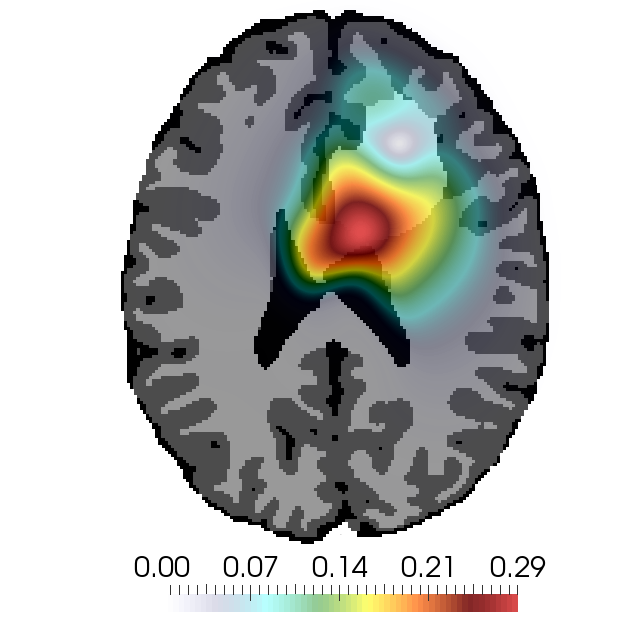}
		\caption{$t = 0.6$}
	\end{subfigure}%
	\begin{subfigure}{.33\textwidth}
		\centering
		\includegraphics[height = 1.\linewidth,width=1.\linewidth]{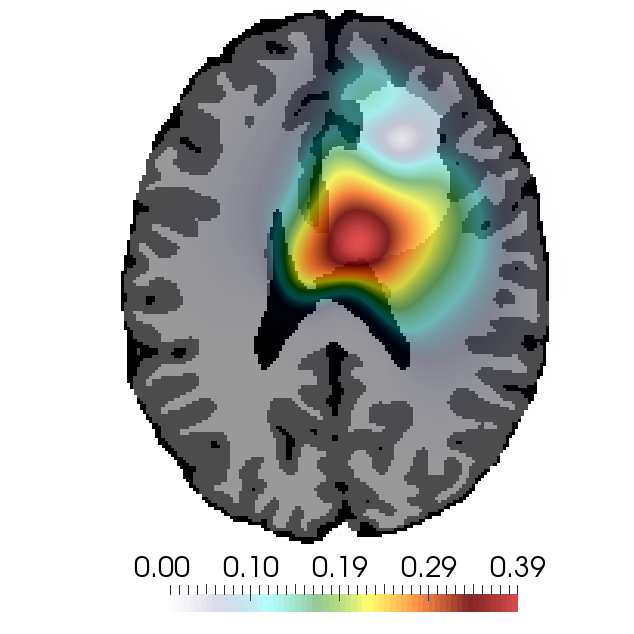}
		\caption{$t = 75$}
	\end{subfigure}
	\caption{Time evolution of the point-wise $l_2$ norm of the displacement field, $u$.}
	\label{f:rd_disp}
\end{figure}

\textit{Single species mass effect: }We show an exemplary simulation for the single species model in~\fref{f:rd_seg}. As we can see, there
is a significant mass effect due to the growth of tumor on the surrounding tissues. The corresponding
point-wise $l_2$ norm of the displacement fields are also highlighted in~\fref{f:rd_disp}. Tissue deformation can be observed
around the tumor boundary and it increases as the tumor grows and spreads. Specifically, we observe large displacements
around the cerebrospinal fluid due to its soft and highly compressible nature. 

\begin{figure}[!htbp]
	\begin{subfigure}{.33\textwidth}
		\centering
		\includegraphics[height = 0.9\linewidth,width=0.9\linewidth]{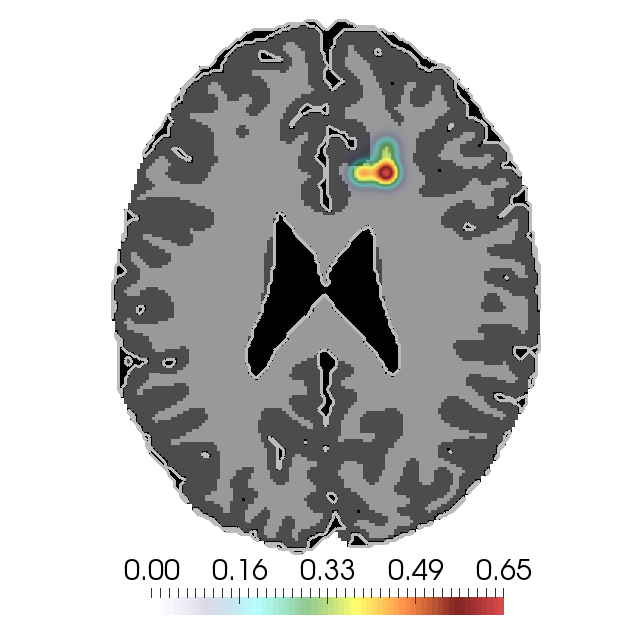}
	\end{subfigure}%
	\begin{subfigure}{.33\textwidth}
		\centering
		\includegraphics[height = 0.9\linewidth,width=0.9\linewidth]{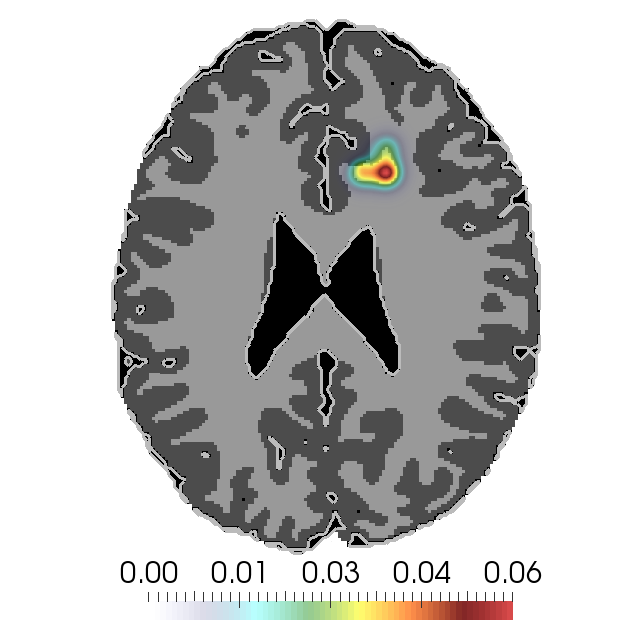}
	\end{subfigure}%
	\begin{subfigure}{.33\textwidth}
		\centering
		\includegraphics[height = 0.9\linewidth,width=0.9\linewidth]{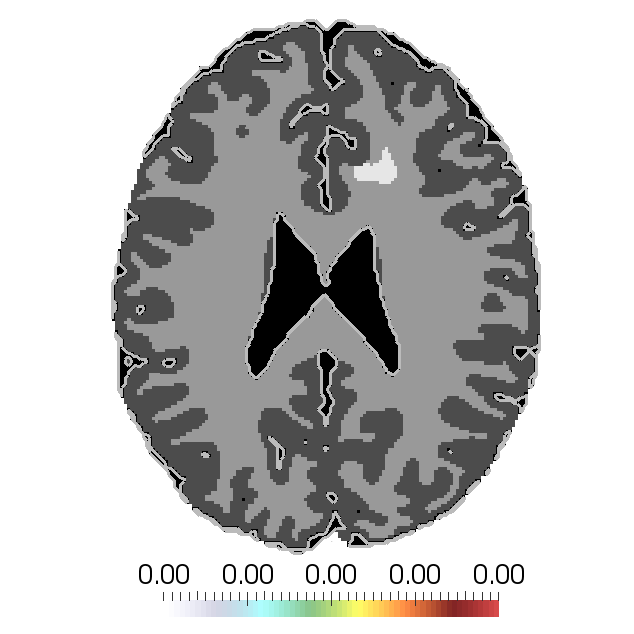}
	\end{subfigure}
	
	\begin{subfigure}{.33\textwidth}
		\centering
		\includegraphics[height = 0.9\linewidth,width=0.9\linewidth]{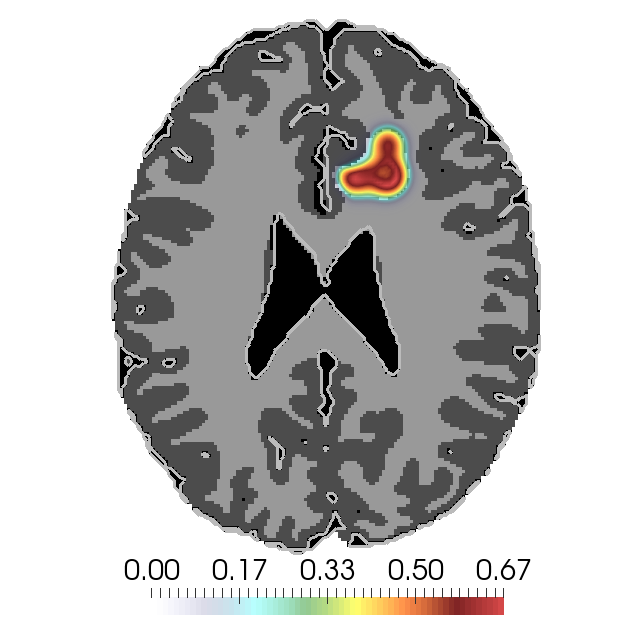}
	\end{subfigure}%
	\begin{subfigure}{.33\textwidth}
		\centering
		\includegraphics[height = 0.9\linewidth,width=0.9\linewidth]{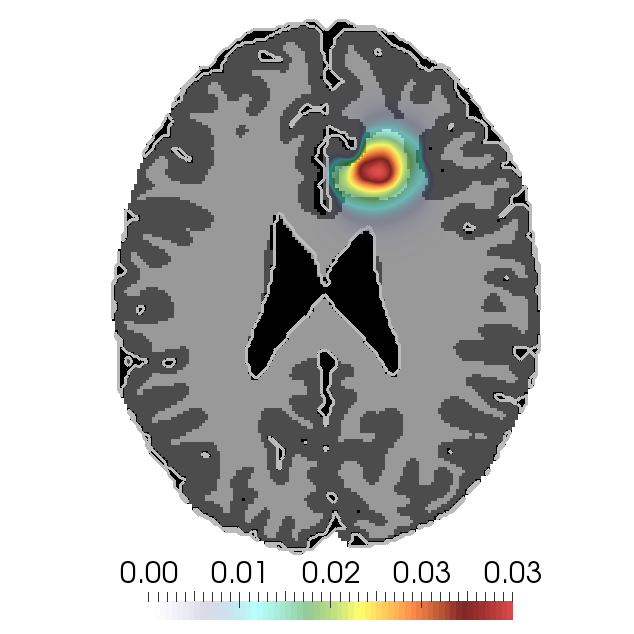}
	\end{subfigure}%
	\begin{subfigure}{.33\textwidth}
		\centering
		\includegraphics[height = 0.9\linewidth,width=0.9\linewidth]{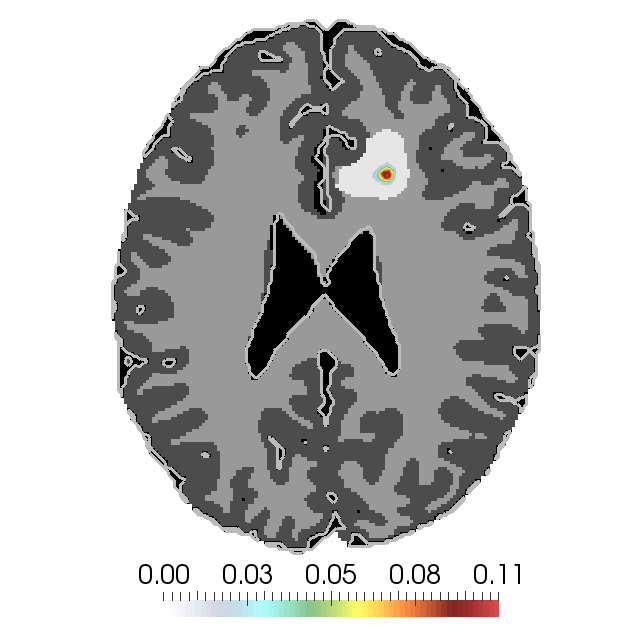}
	\end{subfigure}
	
	\begin{subfigure}{.33\textwidth}
		\centering
		\includegraphics[height = 0.9\linewidth,width=0.9\linewidth]{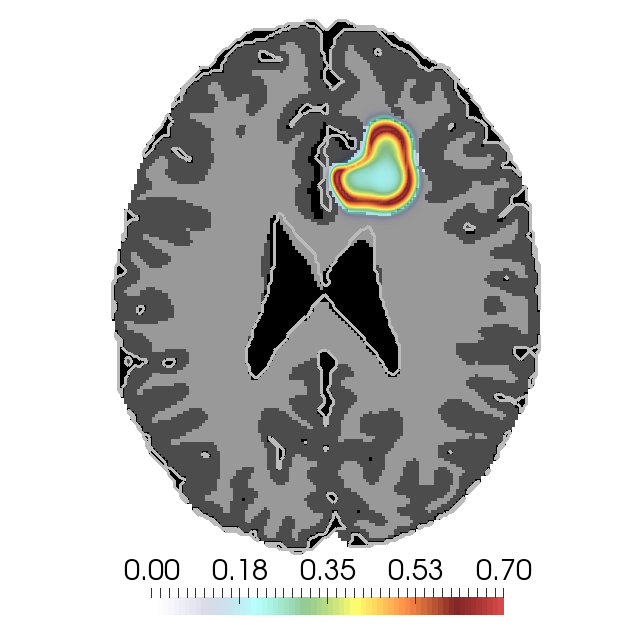}
	\end{subfigure}%
	\begin{subfigure}{.33\textwidth}
		\centering
		\includegraphics[height = 0.9\linewidth,width=0.9\linewidth]{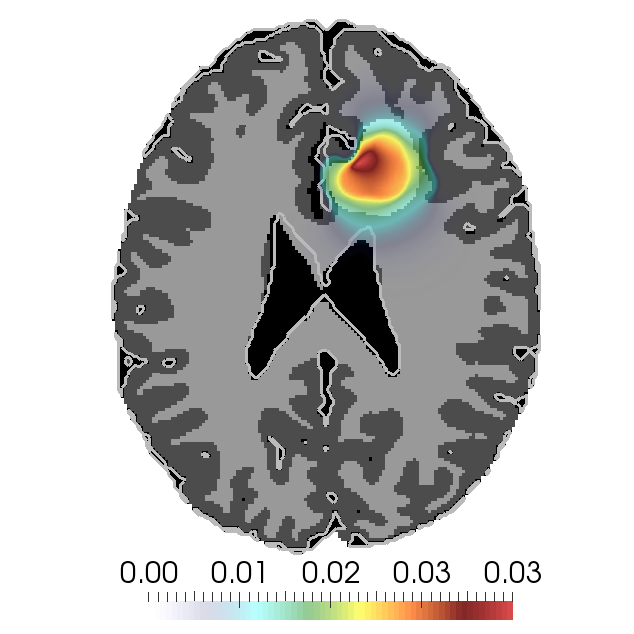}
	\end{subfigure}%
	\begin{subfigure}{.33\textwidth}
		\centering
		\includegraphics[height = 0.9\linewidth,width=0.9\linewidth]{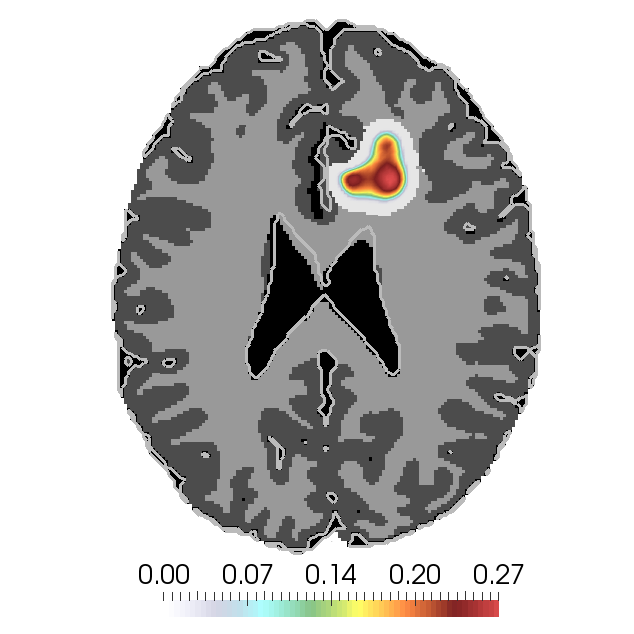}
	\end{subfigure}
	
	\begin{subfigure}{.33\textwidth}
		\centering
		\includegraphics[height = 0.9\linewidth,width=0.9\linewidth]{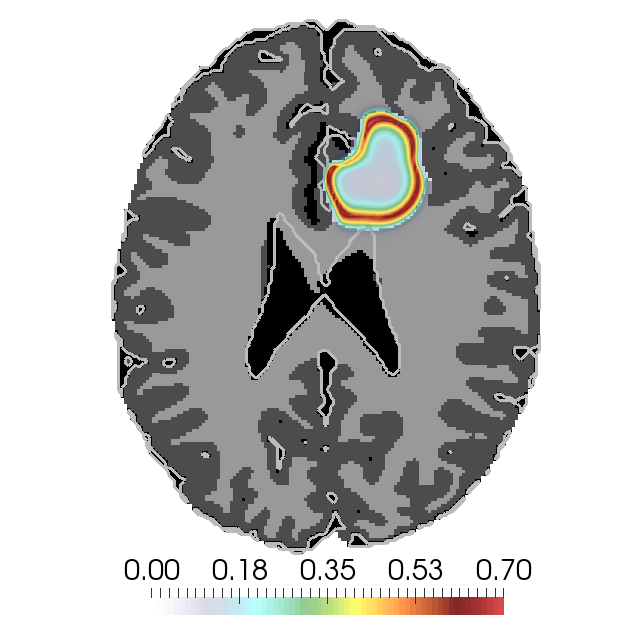}
	\end{subfigure}%
	\begin{subfigure}{.33\textwidth}
		\centering
		\includegraphics[height = 0.9\linewidth,width=0.9\linewidth]{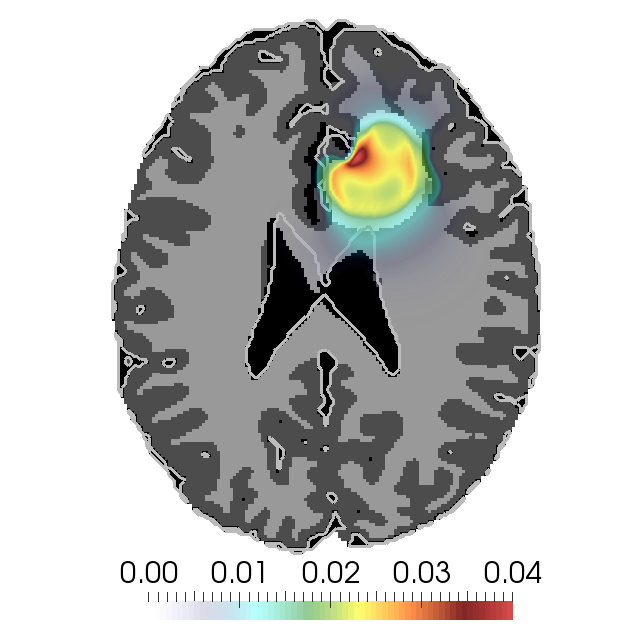}
	\end{subfigure}%
	\begin{subfigure}{.33\textwidth}
		\centering
		\includegraphics[height = 0.9\linewidth,width=0.9\linewidth]{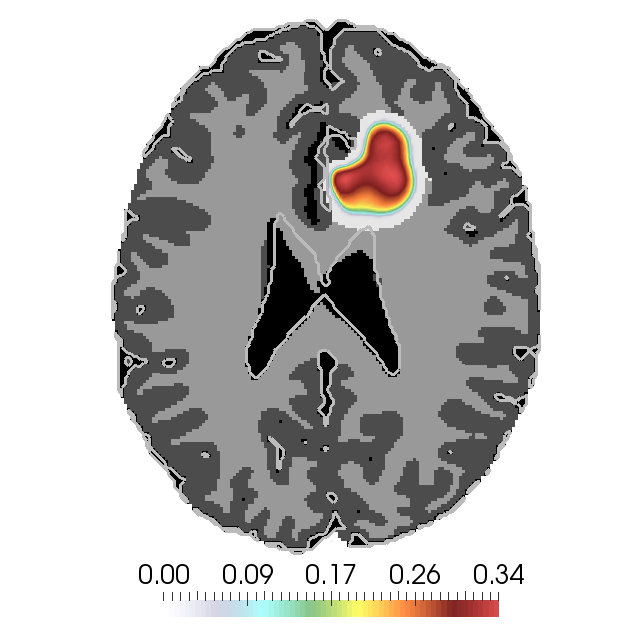}
	\end{subfigure}
	
	\begin{subfigure}{.33\textwidth}
		\centering
		\includegraphics[height = 0.9\linewidth,width=0.9\linewidth]{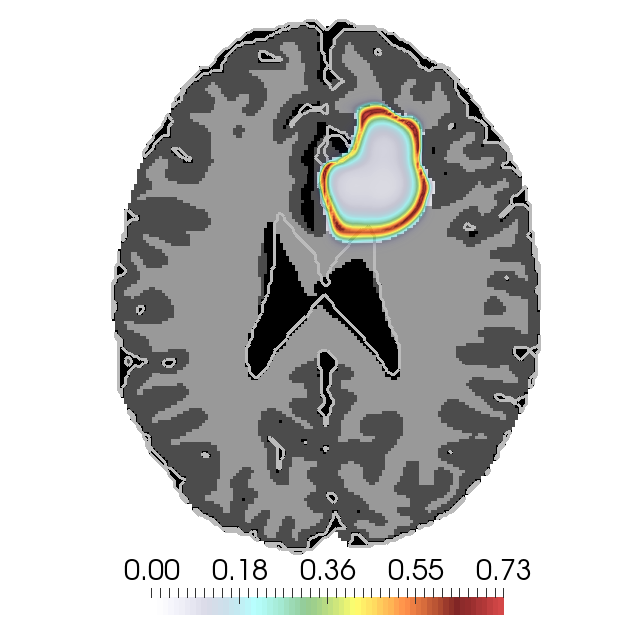}
	\end{subfigure}%
	\begin{subfigure}{.33\textwidth}
		\centering
		\includegraphics[height = 0.9\linewidth,width=0.9\linewidth]{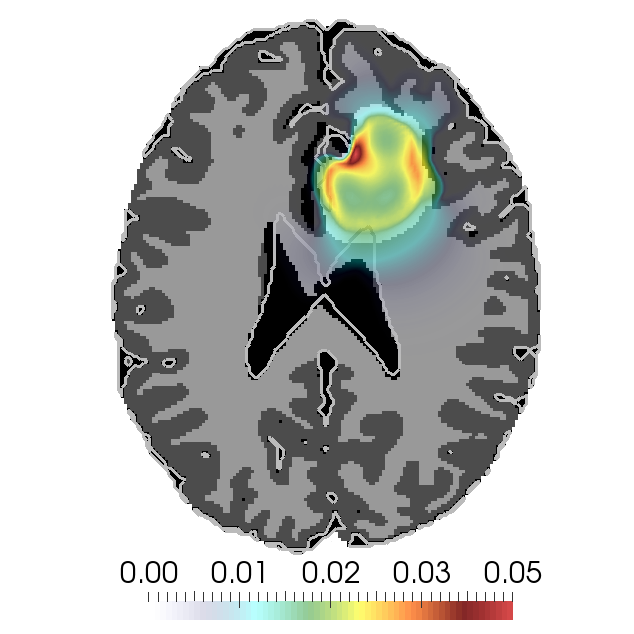}
	\end{subfigure}%
	\begin{subfigure}{.33\textwidth}
		\centering
		\includegraphics[height = 0.9\linewidth,width=0.9\linewidth]{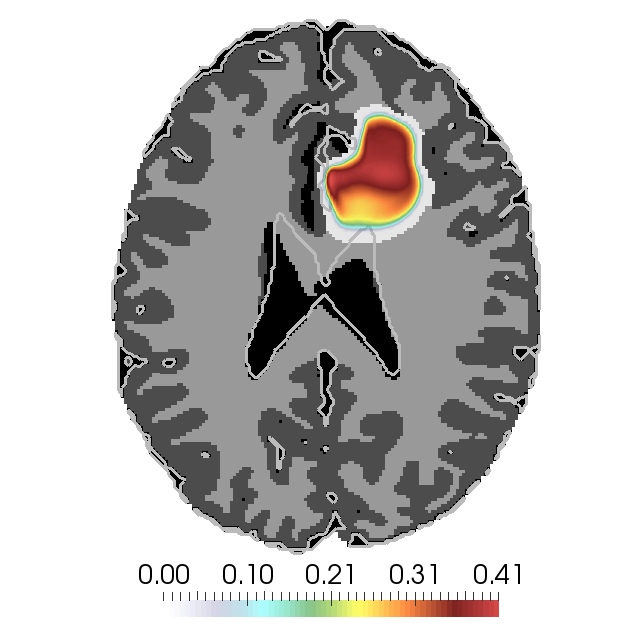}
	\end{subfigure}
	
	\begin{subfigure}{.33\textwidth}
		\centering
		\includegraphics[height = 0.9\linewidth,width=0.9\linewidth]{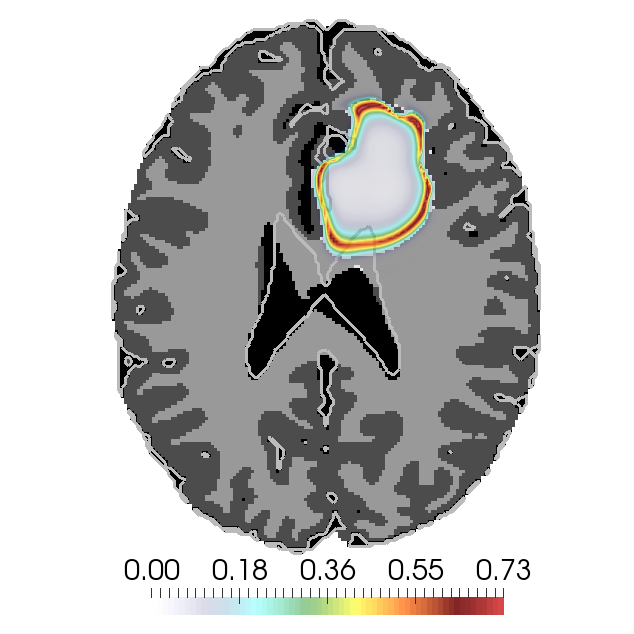}
	\end{subfigure}%
	\begin{subfigure}{.33\textwidth}
		\centering
		\includegraphics[height = 0.9\linewidth,width=0.9\linewidth]{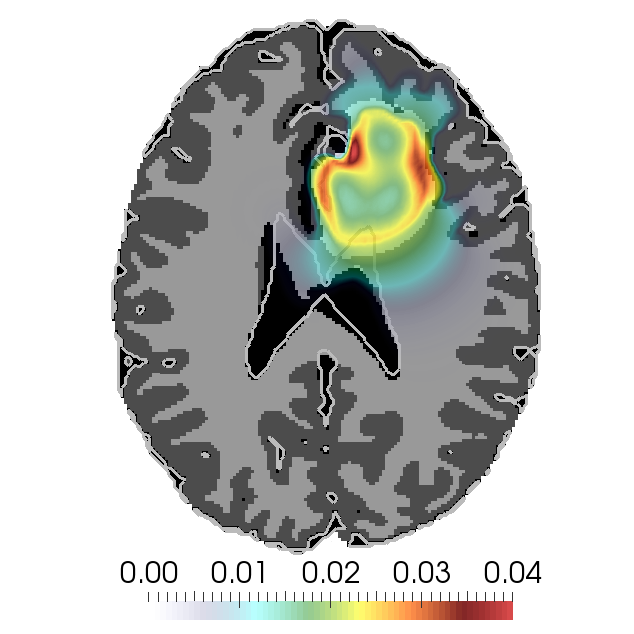}
	\end{subfigure}%
	\begin{subfigure}{.33\textwidth}
		\centering
		\includegraphics[height = 0.9\linewidth,width=0.9\linewidth]{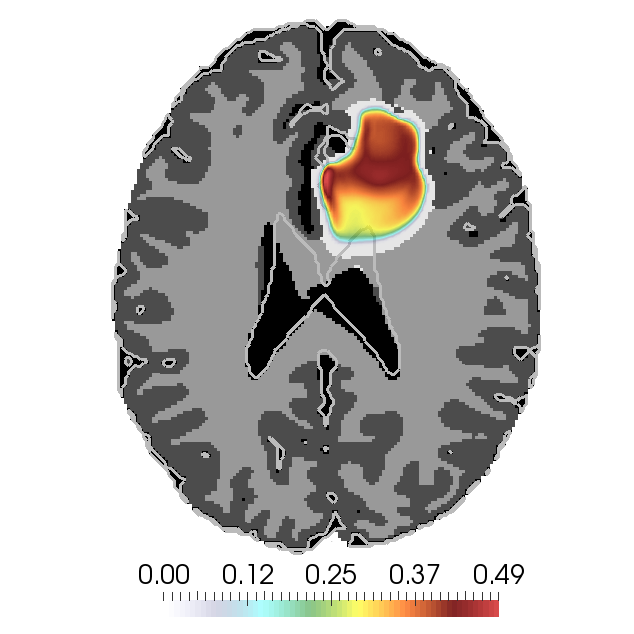}
	\end{subfigure}
	\caption{Time evolution of proliferative (\textit{column 1}), infiltrative (\textit{column 2}), necrotic (\textit{column 3}) tumor cell concentrations. The concentrations are overlayed on the whole tumor core segmentation (proliferative and necrotic tumor). The rows show different time instances of the simulation: $t = 0, 0.2, 0.4, 0.6, 0.8, 1.0$.}
	\label{f:gog_seg}
\end{figure}

\textit{Multispecies mass effect: } The results from the multispecies model with mass effect are shown in~\fref{f:gog_seg}. The initial condition for proliferative cells is a Gaussian mixture and the invasive cells are taken as a small fraction of initial proliferative cell concentration. To better visualize the evolution of the different tumor cell types, we count all tumor phenotypes as one for the segmentation and overlay individual tumor concentrations. 

\begin{figure}[!htbp]
	\begin{subfigure}{.33\textwidth}
		\centering
		\includegraphics[height = 1\linewidth,width=1\linewidth]{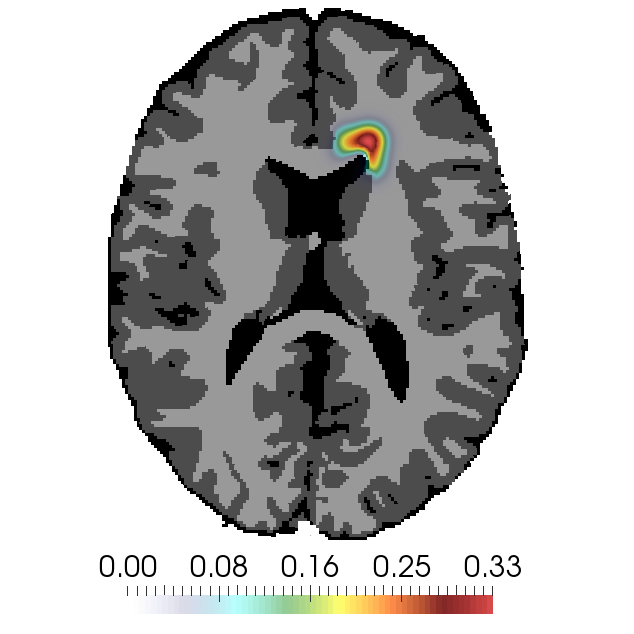}
	\end{subfigure}%
	\begin{subfigure}{.33\textwidth}
		\centering
		\includegraphics[height = 1\linewidth,width=1\linewidth]{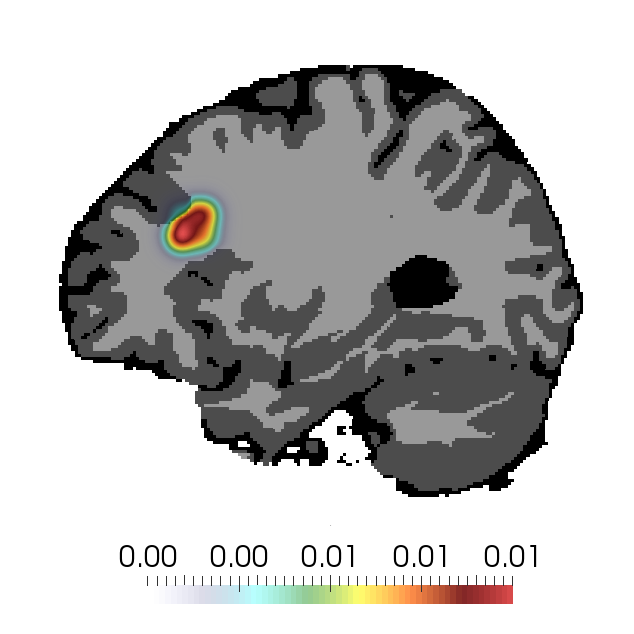}
	\end{subfigure}%
	\begin{subfigure}{.33\textwidth}
		\centering
		\includegraphics[height = 1\linewidth,width=1\linewidth]{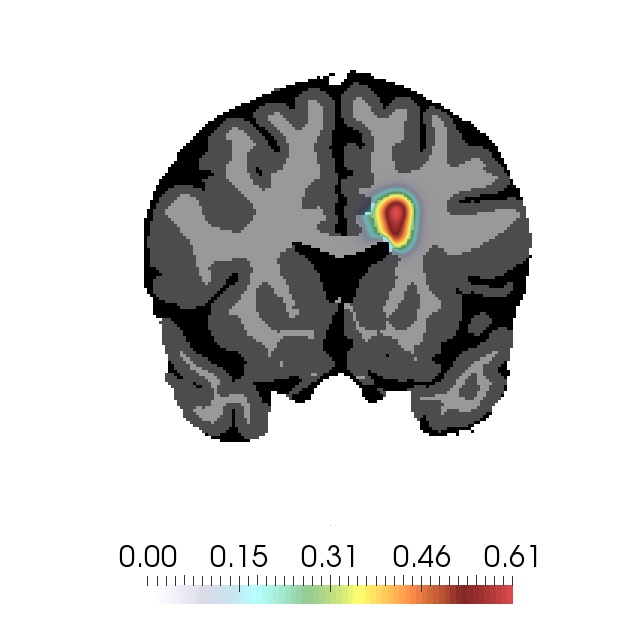}
	\end{subfigure}
	\begin{subfigure}{.33\textwidth}
		\centering
		\includegraphics[height = 1\linewidth,width=1\linewidth]{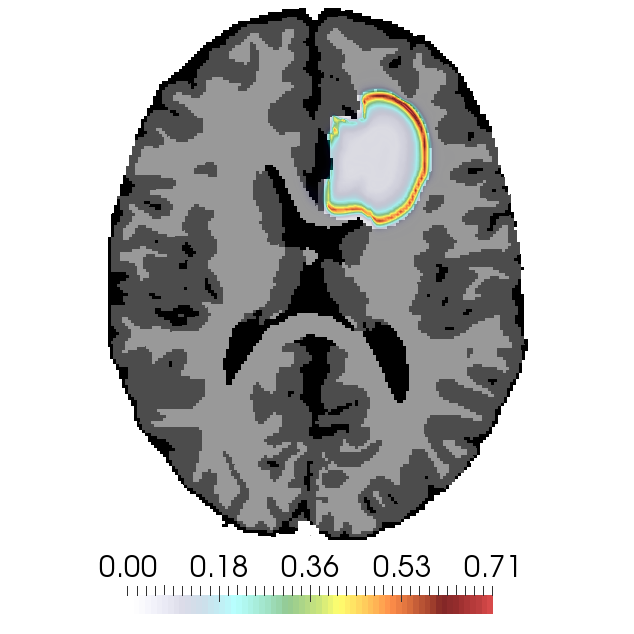}
	\end{subfigure}%
	\begin{subfigure}{.33\textwidth}
		\centering
		\includegraphics[height = 1\linewidth,width=1\linewidth]{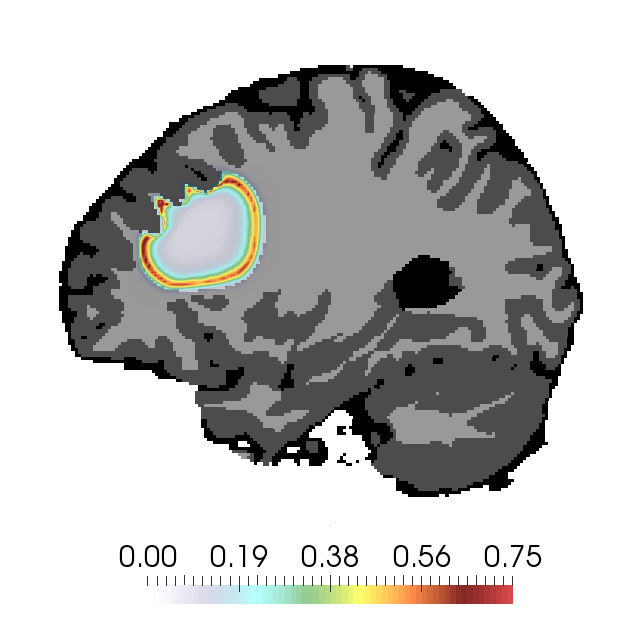}
	\end{subfigure}%
	\begin{subfigure}{.33\textwidth}
		\centering
		\includegraphics[height = 1\linewidth,width=1\linewidth]{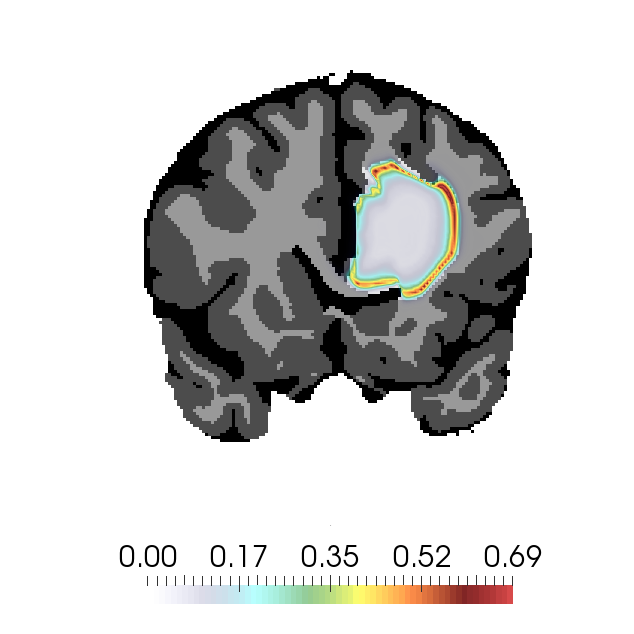}
	\end{subfigure}%
	\caption{Segmented images of 3D simulation showing the evolution of proliferating cells using the multispecies mass effect model at $t = 0$ (\textit{top row}) and $t = 1$ (
		\textit{bottom row}) for axial slice 123, sagittal slice 97 and coronal slice 161.}
	\label{f:3d_rd}
\end{figure}

We can see the characteristic multicomponent structure of a glioblastoma along with significant mass effect on the surrounding tissues in these simulations. This structure includes the necrotic tumor and tissue cells accumulating in a central core surrounded by an expanding rim of proliferating tumor cells. Regarding invasive tumor cells, recent histopathology reports ~\citep{eidel2017histopathology, gill2014} show that they infiltrate or diffuse to regions beyond the enhancing cancer rim. They also indicate that the invasive cell density is maximum in or around a central necrotic tumor region and becomes smaller as we move away from this region. We can observe these trends in invasive cell concentrations in our simulations. A 3D simulation of multispecies mass effect is shown in~\fref{f:3d_rd}. The spatial discretization is equal in all directions. 

\begin{figure}[!htbp]
	\begin{subfigure}{.33\textwidth}
		\centering
		\includegraphics[height = 1\linewidth,width=1\linewidth]{p_brats_t100}
		\caption{Proliferating tumor}
	\end{subfigure}%
	\begin{subfigure}{.33\textwidth}
		\centering
		\includegraphics[height = 1.\linewidth,width=1\linewidth]{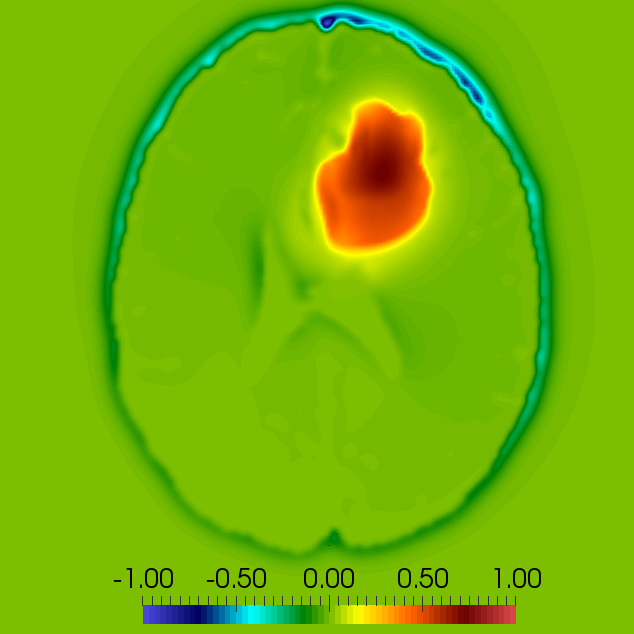}
		\label{f:stress_principal}
		\caption{Trace of stress tensor}
	\end{subfigure}%
	\begin{subfigure}{.33\textwidth}
		\centering
		\includegraphics[height = 1.\linewidth,width=1\linewidth]{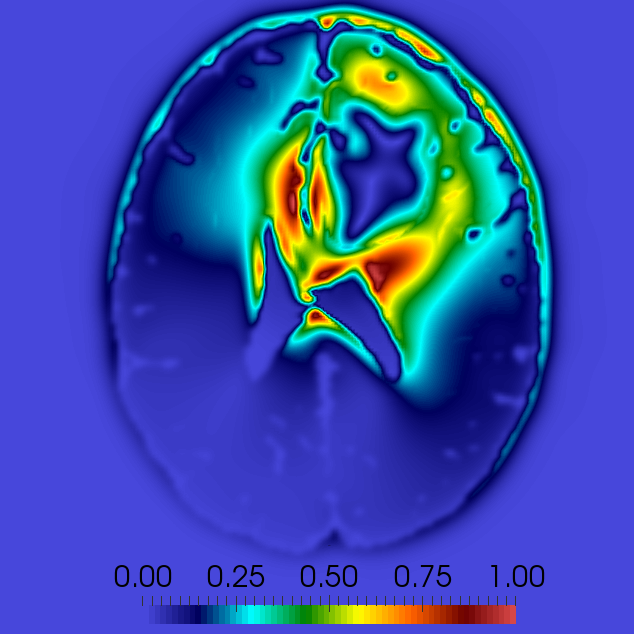}
		\label{f:stress_max_shear}
		\caption{Maximum shear stress}
	\end{subfigure}%
	\caption{Normalized stress plots for single species mass effect at $t = 1$.}
	\label{f:stress}
\end{figure}

\textit{Stress fields: }In order to visualize the stress fields, we show point-wise values for
the principal stresses and maximum shear stress in ~\fref{f:stress}. We outline how these are computed from the
stress tensor in~\eref{e:princ_stress}. The principal stress plot shows small tensile stresses induced
in the tumor core and larger compressive stresses around the tumor boundary. The maximum shear stress is induced around
the growing tumor region. 
\

\textit{Screening effects: }We perform simulations for three different values of the screening coefficient outside the tumor ($\eta_{\text{healthy cells}}$) to observe its effects.

\begin{figure}[!htbp]
	\begin{subfigure}{.33\textwidth}
		\centering
		\includegraphics[height = 1\linewidth,width=1\linewidth]{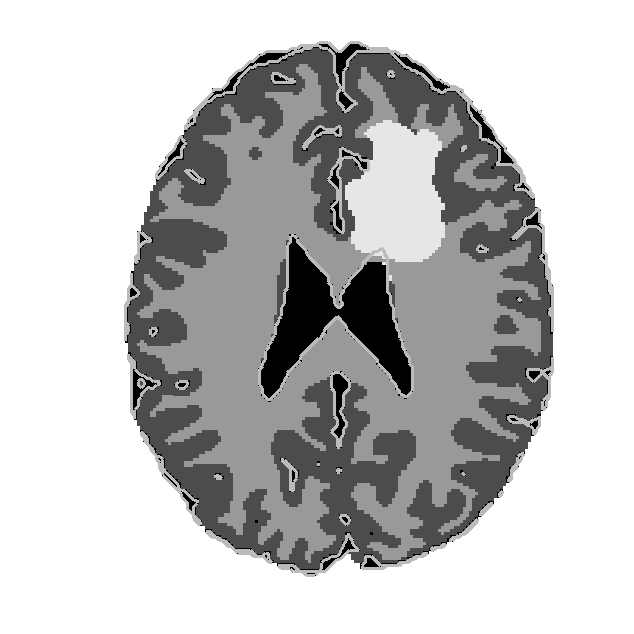}
		\caption{$\eta_{\text{healthy cells}} = 10^6$}
		\label{f:seg_comp_a}
	\end{subfigure}%
	\begin{subfigure}{.33\textwidth}
		\centering
		\includegraphics[height = 1\linewidth,width=1\linewidth]{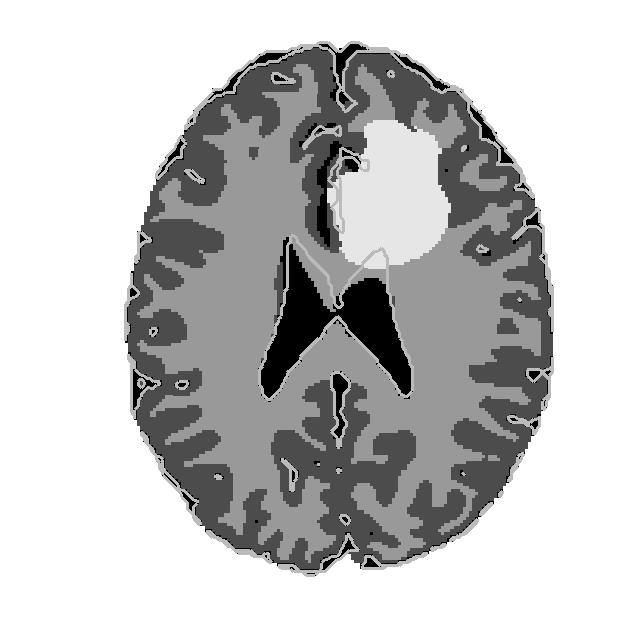}
		\caption{$\eta_{\text{healthy cells}} = 10^3$}
		\label{f:seg_comp_b}
	\end{subfigure}%
	\begin{subfigure}{.33\textwidth}
		\centering
		\includegraphics[height = 1\linewidth,width=1\linewidth]{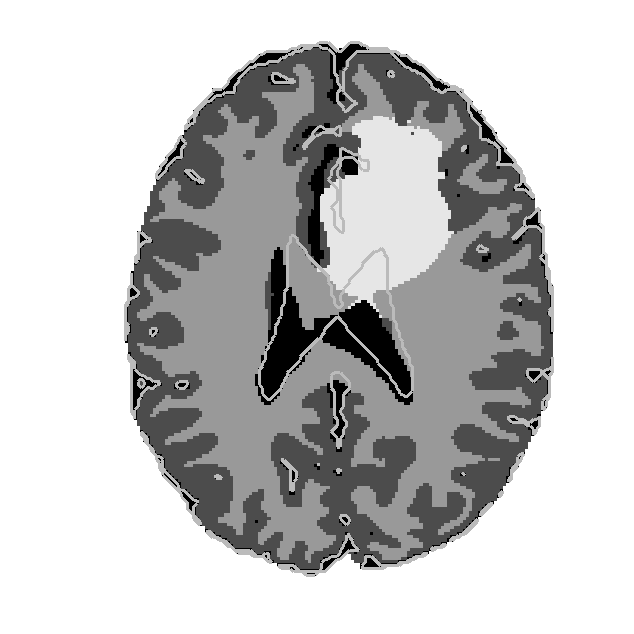}
		\caption{$\eta_{\text{healthy cells}} = 0$}
		\label{f:seg_comp_c}
	\end{subfigure}%
	\caption{Effect of screening coefficient, $\eta_{\text{healthy cells}}$ on cerebrospinal fluid deformation. The cerebrospinal fluid contours are overlayed on the initial segmentation for three screening coefficients.
	}
	\label{f:seg_comp}
\end{figure}

We visualize this in~\fref{f:seg_comp} using contours of cerebrospinal fluid to indicate the extent of its
deformation for the three cases. A large value of $\eta_{\text{healthycells}}$ indicates a highly localized mass effect to regions immediately around the tumor core. Smaller values result in observable mass effect in regions far away from the tumor core. However, for tumors close to the ventricles, a long-range mass effect can lead to their excessive shearing. Hence, a reasonable screening parameter is helpful for capturing realistic deformations.

\begin{figure}[!htbp]
	\begin{subfigure}{0.25\linewidth}
		\centering
		\includegraphics[height=1\linewidth, width=1\linewidth]{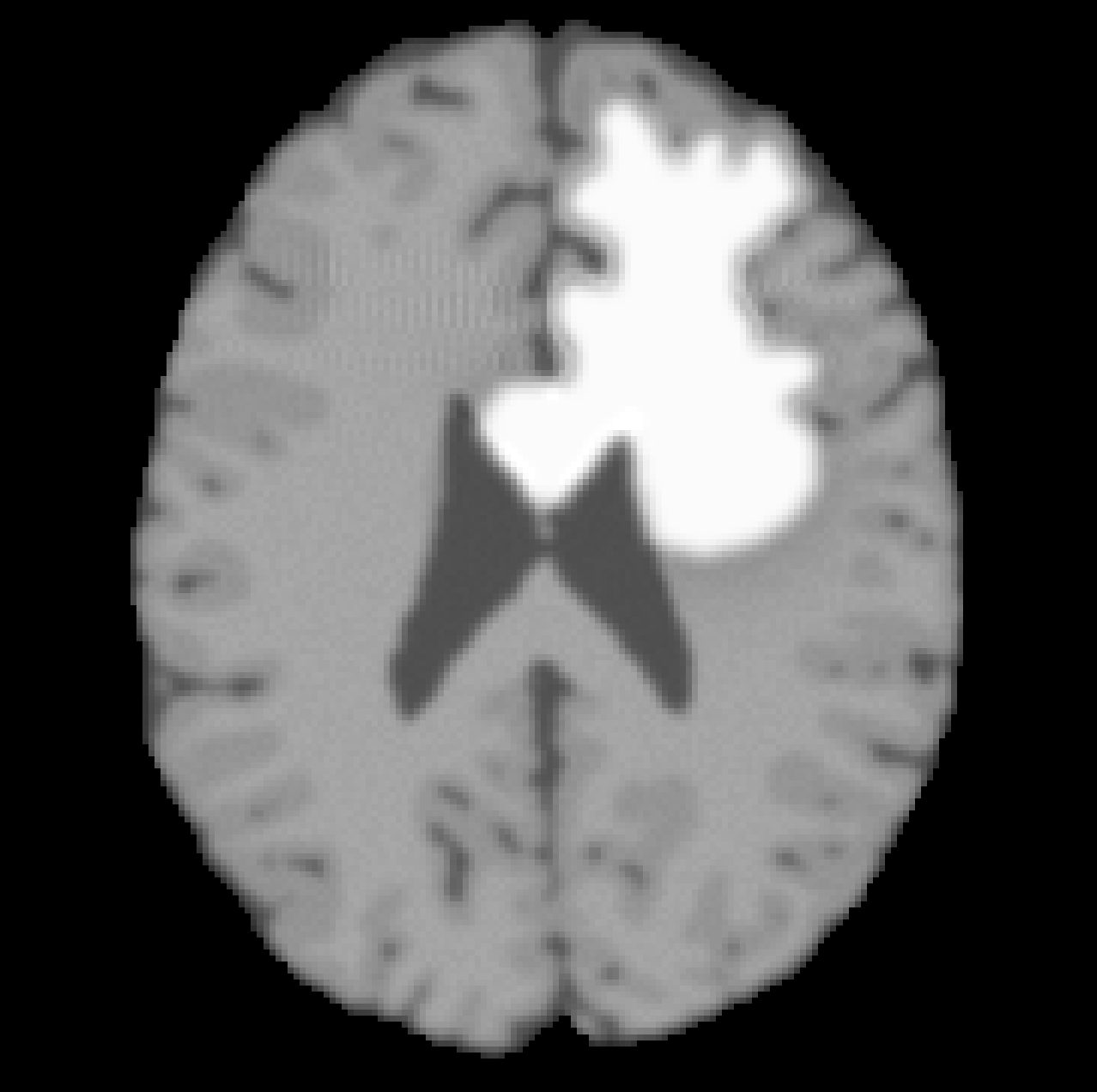}
	\end{subfigure}%
	\begin{subfigure}{0.25\linewidth}
		\centering
		\includegraphics[height=1\linewidth, width=1\linewidth]{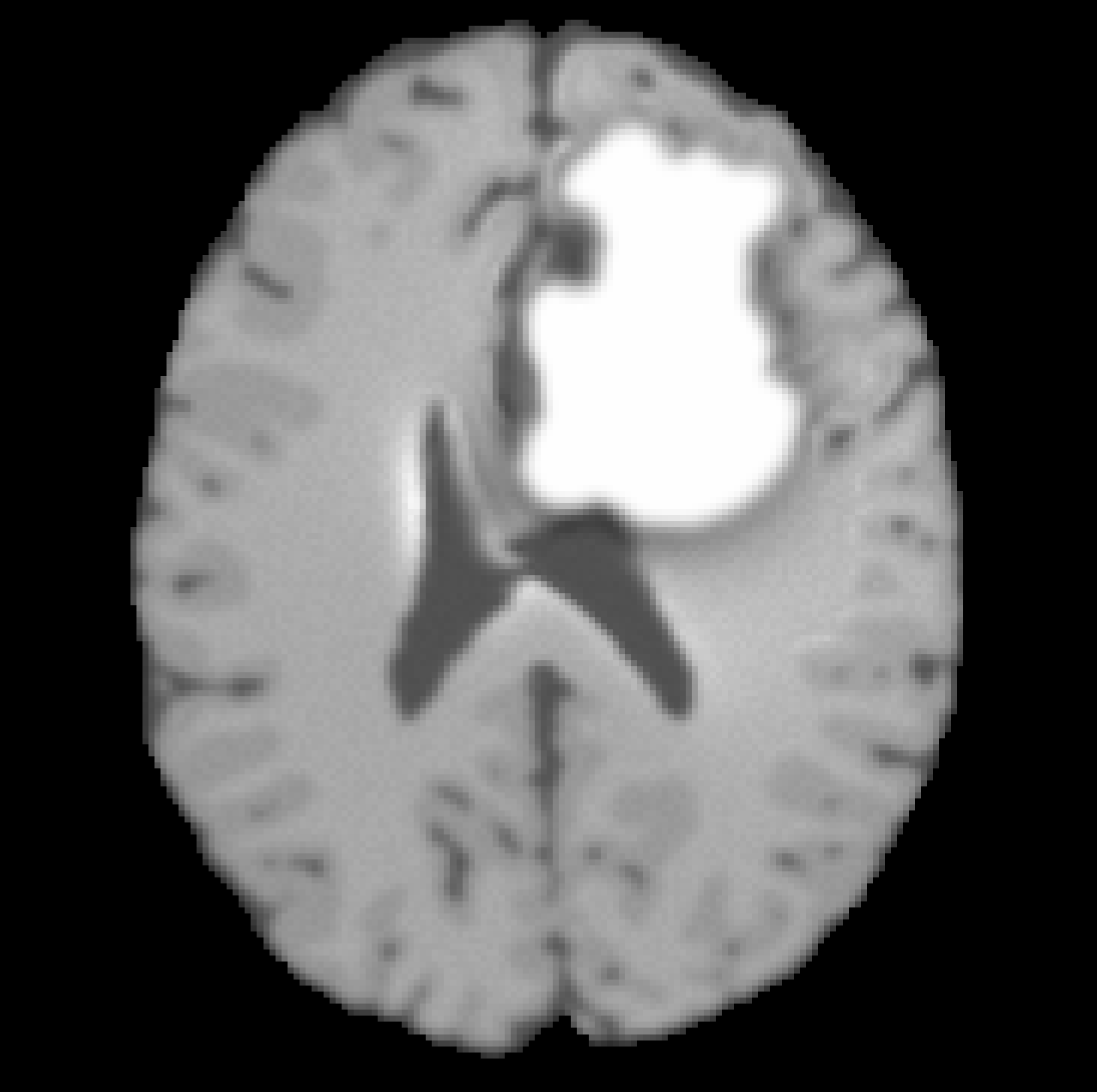}
	\end{subfigure}%
	\begin{subfigure}{0.25\linewidth}
		\centering
		\includegraphics[height=1\linewidth, width=1\linewidth]{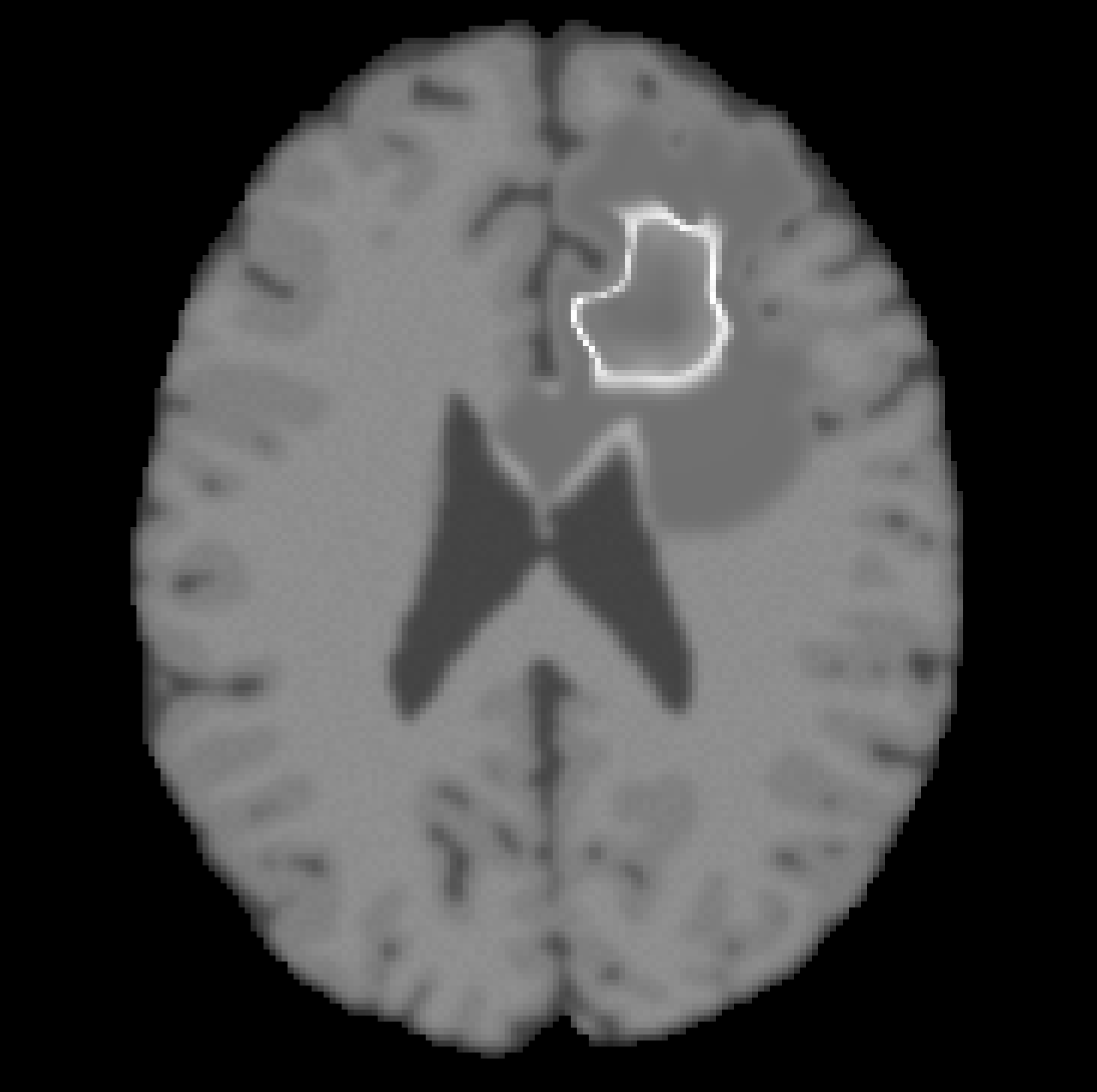}
	\end{subfigure}%
	\begin{subfigure}{0.25\linewidth}
		\centering
		\includegraphics[height=1\linewidth, width=1\linewidth]{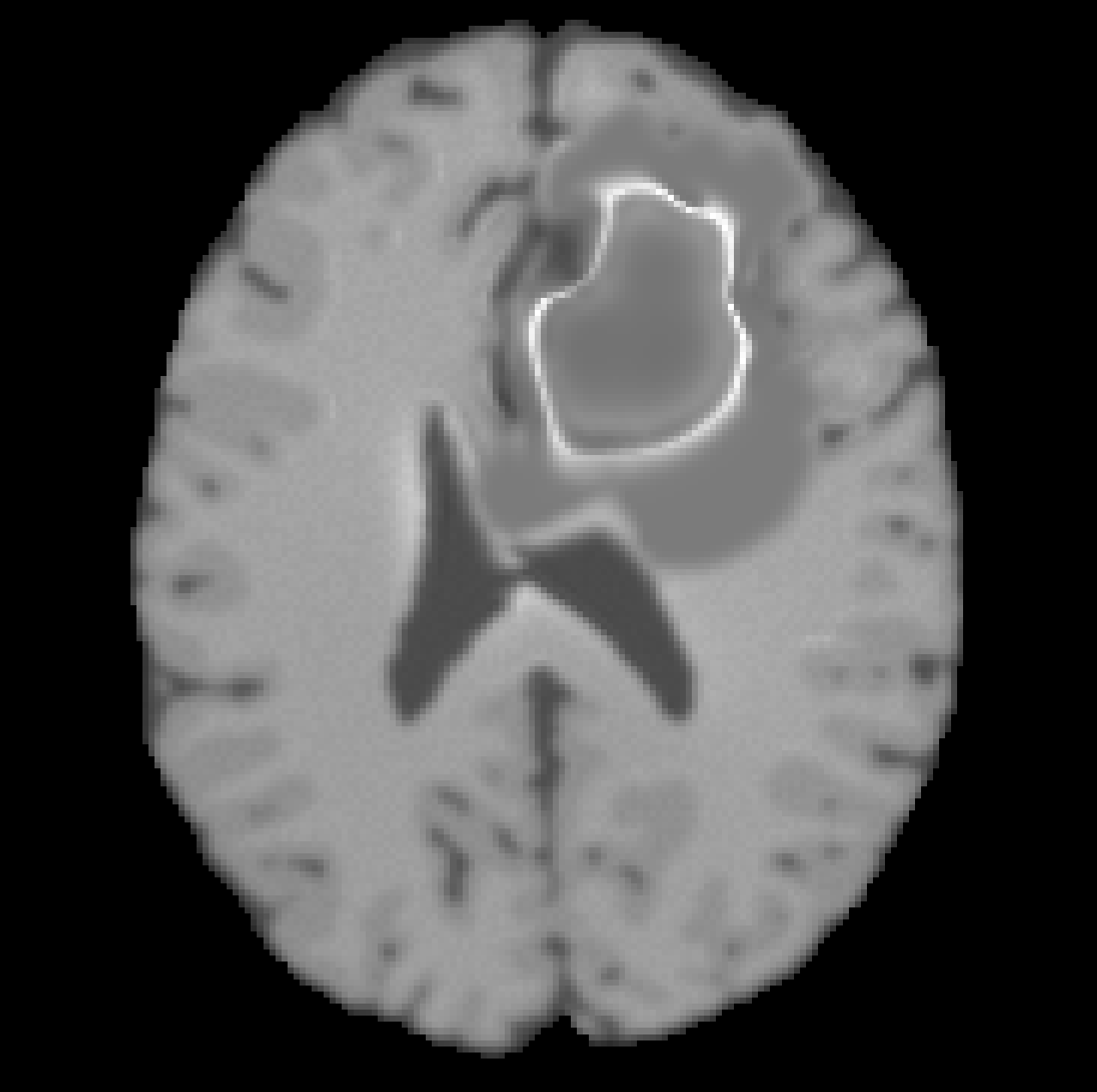}
	\end{subfigure}%
	\caption{Simulated T1-Gd MRI images: \textit{(left to right)} Single species model with and without mass effect, multispecies model with and without mass effect. The intensities for the tumor cells in the single species model are sampled from enhancing tumor structures of real MRI scans.}
	\label{f:mri_comp}
\end{figure}

\begin{figure}[!htbp]
	\begin{subfigure}{0.25\linewidth}
		\centering
		\includegraphics[height=1\linewidth, width=1\linewidth]{casebrats_flair_crop}
	\end{subfigure}%
	\begin{subfigure}{0.25\linewidth}
		\centering
		\includegraphics[height=1\linewidth, width=1\linewidth]{casebrats_t1_crop}
	\end{subfigure}%
	\begin{subfigure}{0.25\linewidth}
		\centering
		\includegraphics[height=1\linewidth, width=1\linewidth]{casebrats_t1ce_crop}
	\end{subfigure}%
	\begin{subfigure}{0.25\linewidth}
		\centering
		\includegraphics[height=1\linewidth, width=1\linewidth]{casebrats_t2_crop}
	\end{subfigure}
	\begin{subfigure}{0.25\linewidth}
		\centering
		\includegraphics[height=1.\linewidth, width=1\linewidth]{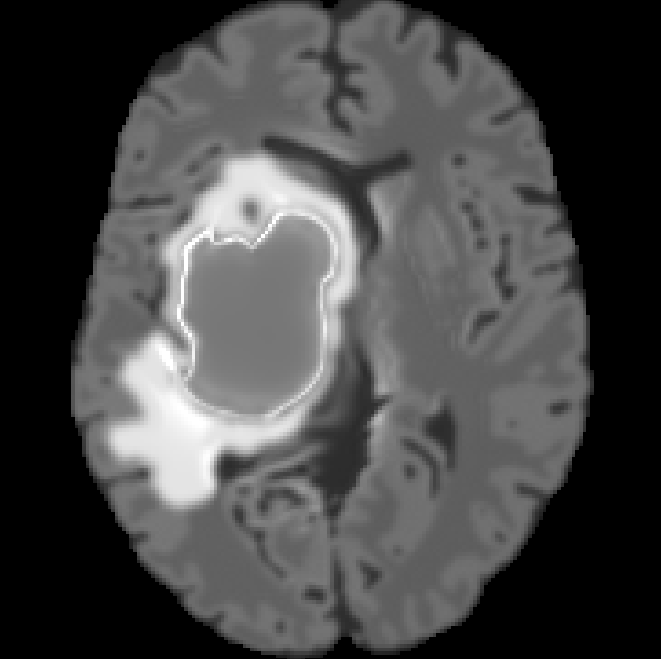}
	\end{subfigure}%
	\begin{subfigure}{0.25\linewidth}
		\centering
		\includegraphics[height=1.\linewidth, width=1\linewidth]{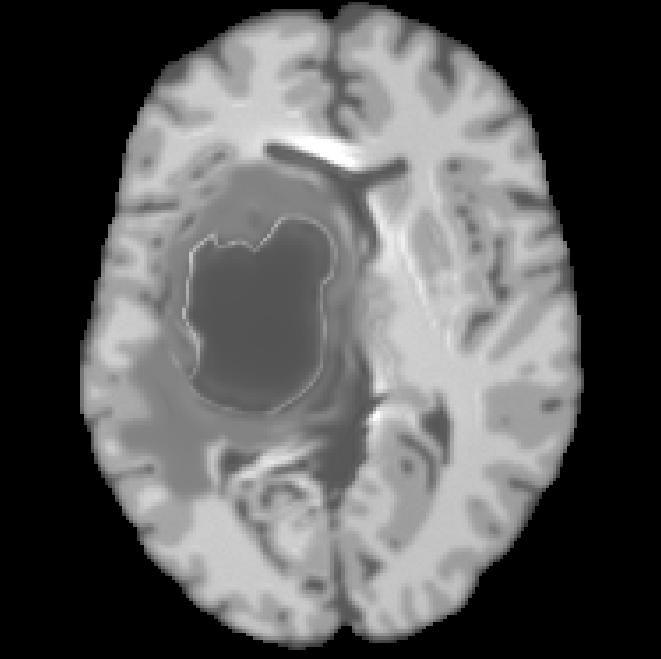}
	\end{subfigure}%
	\begin{subfigure}{0.25\linewidth}
		\centering
		\includegraphics[height=1.\linewidth, width=1\linewidth]{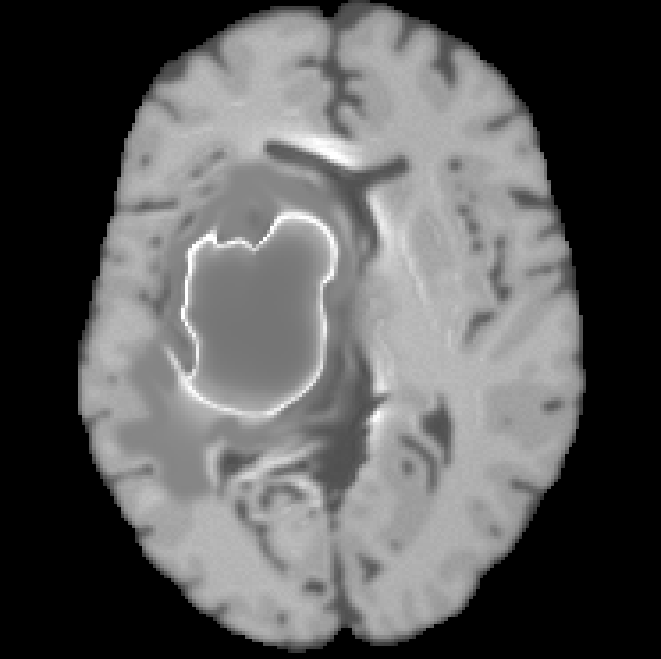}
	\end{subfigure}%
	\begin{subfigure}{0.25\linewidth}
		\centering
		\includegraphics[height=1.\linewidth, width=1\linewidth]{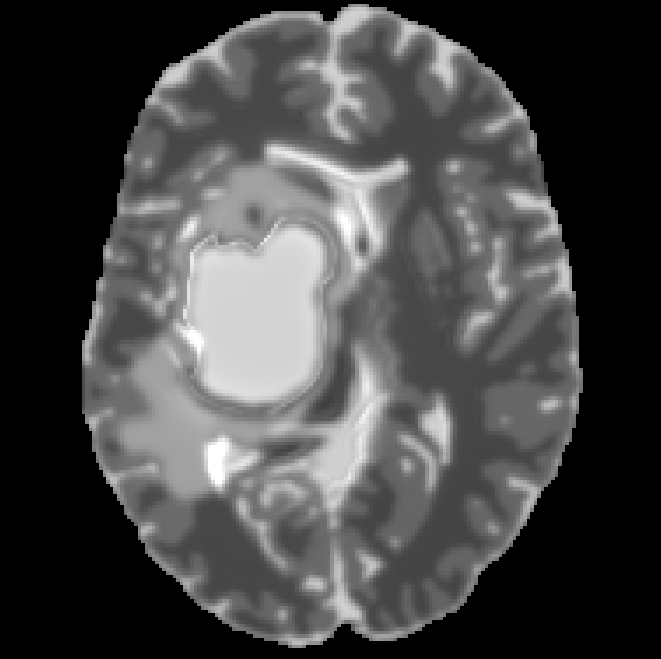}
	\end{subfigure}
	\begin{subfigure}{0.25\linewidth}
		\centering
		\includegraphics[height=1\linewidth, width=1\linewidth]{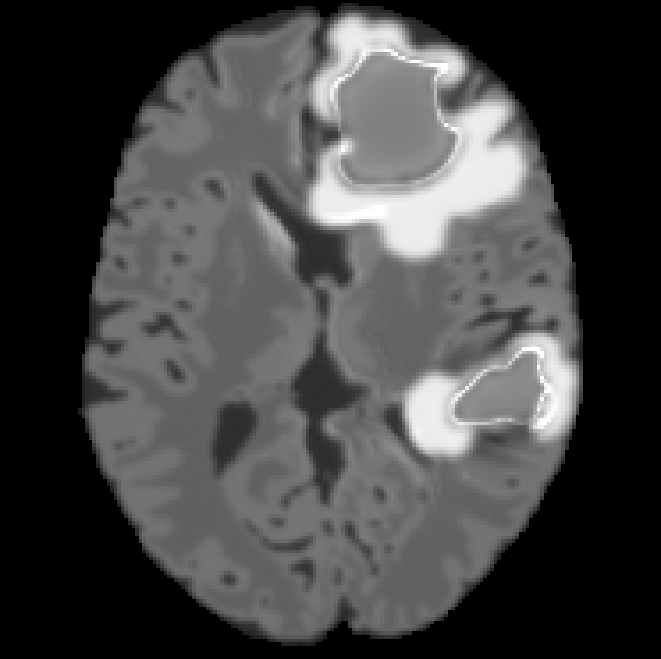}
	\end{subfigure}%
	\begin{subfigure}{0.25\linewidth}
		\centering
		\includegraphics[height=1\linewidth, width=1\linewidth]{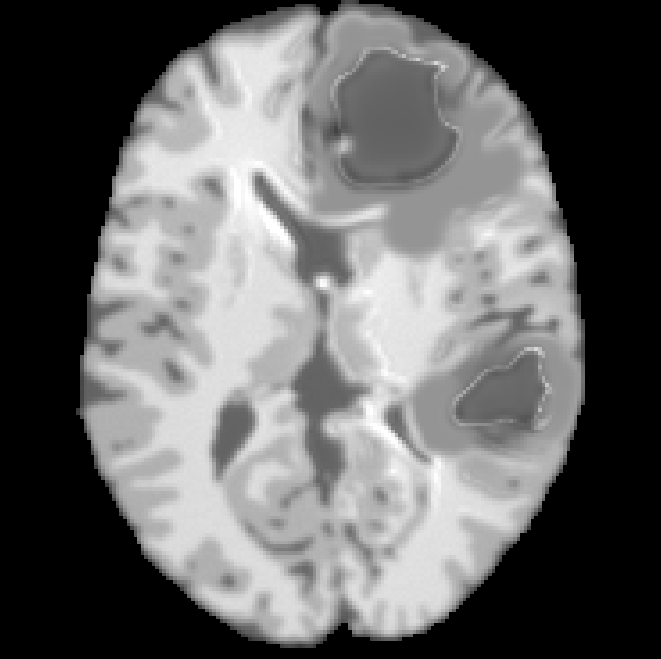}
	\end{subfigure}%
	\begin{subfigure}{0.25\linewidth}
		\centering
		\includegraphics[height=1\linewidth, width=1\linewidth]{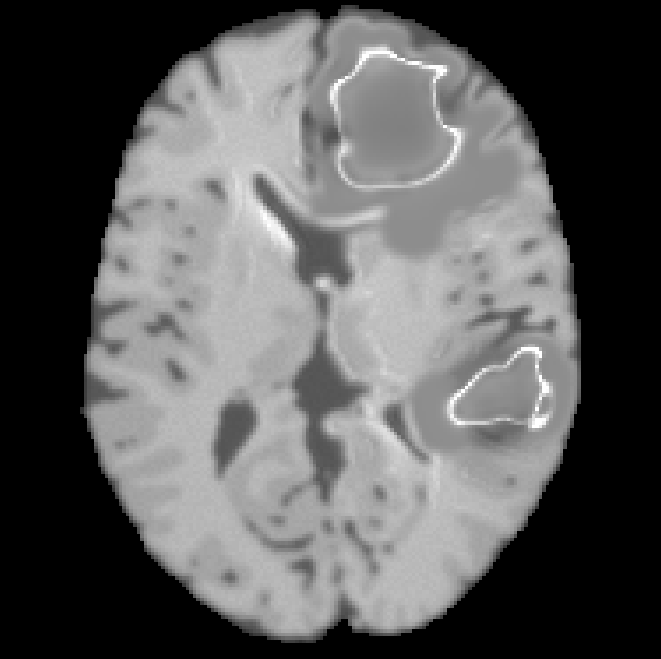}
	\end{subfigure}%
	\begin{subfigure}{0.25\linewidth}
		\centering
		\includegraphics[height=1\linewidth, width=1\linewidth]{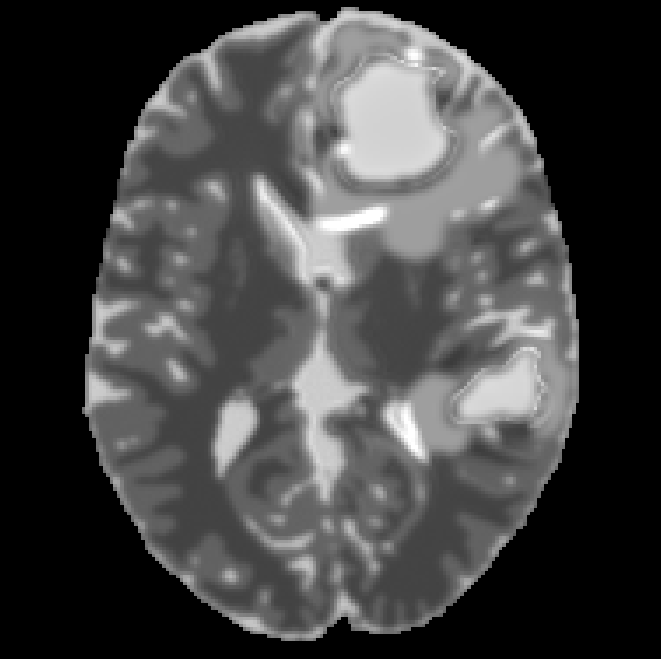}
	\end{subfigure}	
	\caption{Simulated MRI images (FLAIR, T1, T1-Gd and T2) of tumors with different initial conditions. The bottom row shows a multi-focal tumor growth simulation. The hyper-intense regions of the FLAIR images show the extent of edematous fluid and the delineated enhancing and necrotic tumor structures can be seen from the T1-Gd images. All tumors produce significant mass effect on the surrounding brain tissue.}
	\label{f:seg_gog_1}
\end{figure}

\textit{Simulated MRI images: } We use the concentration and segmentation for tumor cells and healthy cells to simulate MRI images by sampling intensities from real MRI scans taken from the GLISTR dataset~\citep{gooya-biros-davatzikos-e12} and the BraTS dataset~\citep{menze2015multimodal}. We compare simulated T1-Gd MRI images from the different models in~\fref{f:mri_comp}. Unlike the multispecies model, the single species model provides no information about the enhancing and necrotic tumor structures. It also does not show edema which can be correlated with the extent of infiltration by tumor cells. Further, without mass effect, neither of the models capture realistic deformations of the ventricles and other tissue types. More exemplar simulated MRI scans are shown in~\fref{f:seg_gog_1}, with varying initial conditions for the tumor location.

\subsection{Sensitivity Analysis}\label{s:sens}
We perform a sensitivity analysis to find which parameters in the model are most sensitive. To determine the effects of the parameters in the multispecies model, we calculate their respective gradients using finite differences, by varying one factor at a time: 
\begin{equation}
g_\iota = \frac{\|\iota\| - \|\iota^*\|}{\epsilon \|\iota^*\|},
\end{equation}
where $\iota$ is the tumor species concentration, $\iota^*$ is an ``optimal" species concentration (obtained using the parameters from~\tref{t:setup}) and $\epsilon$ is a small perturbation to the parameter in consideration. We assess the sensitivity of each tumor species to the different parameters by comparing their respective gradients $g_\iota$. The results are shown in~\tref{t:gog_sensitivity}. Proliferative cell concentrations are most sensitive to the oxygen hypoxia threshold ($o_{\text{hypoxia}}$) and the reaction coefficient ($\rho$). While the reaction coefficient controls the proliferative rate, the hypoxia threshold has strong influence on the conditions which lead to phenotype switching and necrosis. Other oxygen parameters such as the oxygen consumption ($\delta_p$) also affect the concentration, but not as strongly. Invasive cell concentrations are most sensitive to the transition rate from proliferative to invasive cells ($\alpha_0$) and the reaction coefficient since these parameters contribute to the source of invasive cells.  Necrotic cells are affected largely by the death rate ($\gamma$), hypoxia threshold and reaction coefficient. None of the tumor concentrations are sensitive to the transition rate from invasive to proliferative cells ($\beta_0$). The solution doesn't seem to be very sensitive to the diffusion coefficient ($k$) at least for the range of parameters we tested.  These results show a consistent importance of the reaction coefficient and hypoxia threshold for all tumor species. But, the effect of other parameters is variable amongst the different species. 
\begin{table}[!htbp]
	\caption{Sensitivity of the multispecies tumor model on different model parameters evaluated through gradients computed by varying one factor at a time.}
	\label{t:gog_sensitivity}
	\centering
	\begin{tabular}{c|c|c|c}
		\textbf{Parameter} 				&			$g_p$				&     	$g_i$ 		& 			$g_n$ \\
		\hline
		\ghligh
		Reaction coefficient, $\rho$ \hspace{4pt}(see Eq. \eqref{e:reaction_op}) 		& 			0.377				&			1.798			&					0.687		\\
		Hypoxia threshold, $o_{\text{hypoxia}}$	\hspace{4pt}(see Eq. \eqref{e:death})	& 			-0.532				&		-0.124				&			0.369				\\
		\ghligh
		Deathrate, $\gamma$	\hspace{4pt}(see Eq. \eqref{e:death})	& 			-0.166				&		-0.123				&	0.710						\\
		Transition rate from $p$ to $i$, $\alpha_0$	\hspace{4pt}(see Eq. \eqref{e:alpha})	& 		-0.029					&		0.304				&		0					\\
		\ghligh			
		
		Oxygen consumption, $\delta_p$	\hspace{4pt}(see Eq. \eqref{e:ox})	& 		-0.240					&			-0.096			&		0.247					\\			
		Oxygen source, $\delta_s$	\hspace{4pt}(see Eq. \eqref{e:ox})	& 			0.15				&			0.059			&		-0.146		\\
		\ghligh
		Transition rate from $i$ to $p$, $\beta_0$	\hspace{4pt}(see Eq. \eqref{e:beta})		& 			0.003				&		-0.012				&		0					\\
		
		Diffusion coefficient, $k$	\hspace{4pt}(see Eq. \eqref{e:diffusion_tensor})	& 				0.065			&			-0.026			&			-0.042				\\			
		
		\hline			
	\end{tabular}
\end{table}

To assess the sensitivity of the model to mass effect, we perform a grid-search on two parameters: screening coefficient of healthy cells, $\eta_{\text{healthy cells}}$ and forcing function constant, $\zeta$. We report the L2 norm of the Jacobian (determinant of the deformation gradient, see~\eref{e:jacobian}), $\mathcal{J}$ and the maximum distance of a displacement contour (corresponding to a displacement of $0.1$ voxels) from the initial tumor seed, $d_{\text{max}}$. This threshold was chosen to visualize the localization effects of the screening coefficient. 

\begin{figure}
	\centering
	\includegraphics[height=0.5\linewidth, width=.9\linewidth]{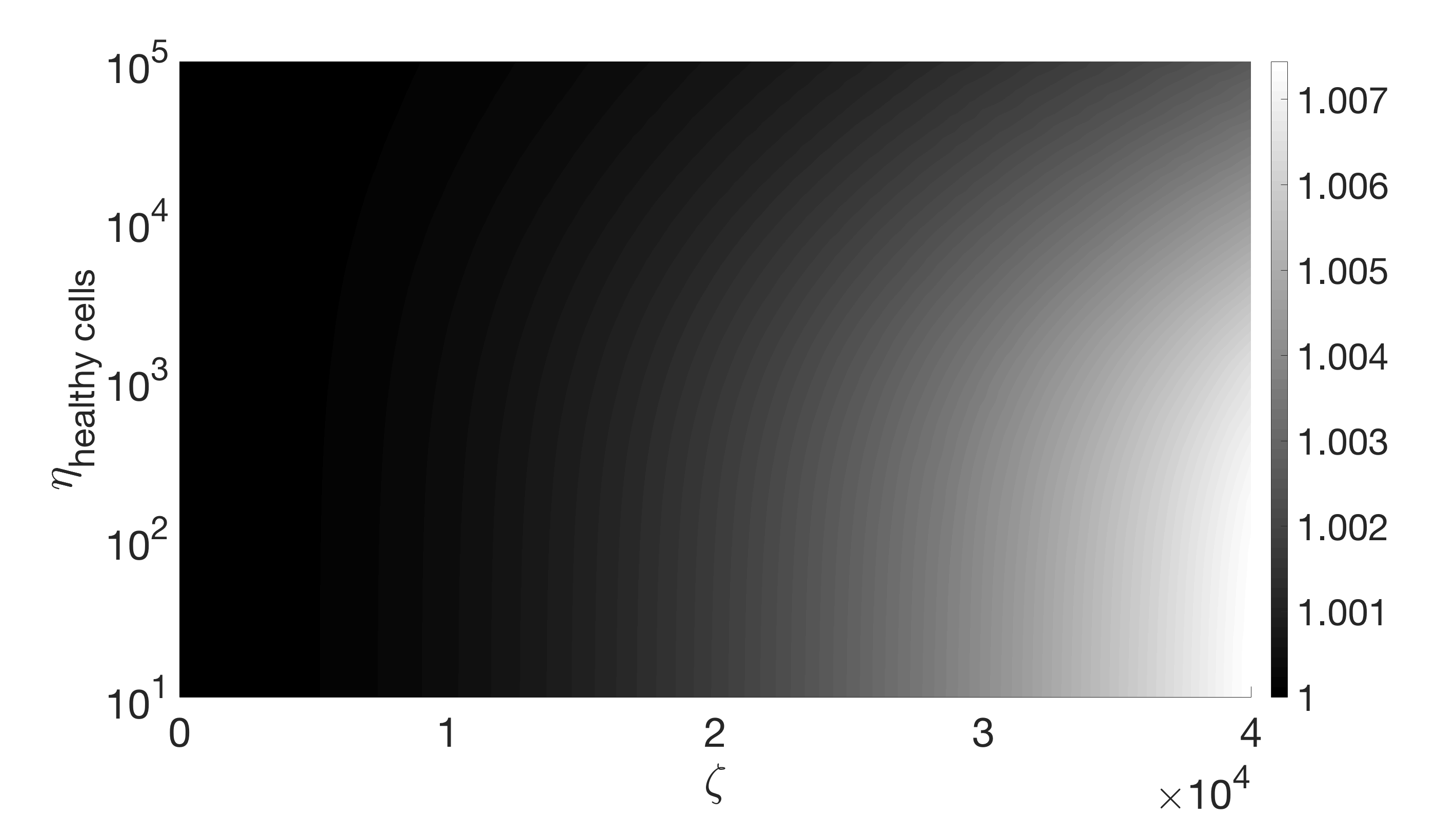}
	\caption{Plot of the L2 norm of the re-scaled deformation Jacobian $\mathcal{J}$ (see~\eref{e:jacobian}). The L2 norm of the Jacobian is large for higher forcing factors $\zeta$ and lower screening coefficients $\eta_{\text{healthy cells}}$. High $\zeta$ values means large body forces and hence larger Jacobians. With smaller screening coefficients, the range mass effect is longer resulting in larger Jacobians.} \label{f:mes_a}
\end{figure}

\begin{figure}
	\centering
	\includegraphics[height=0.5\linewidth, width=.9\linewidth]{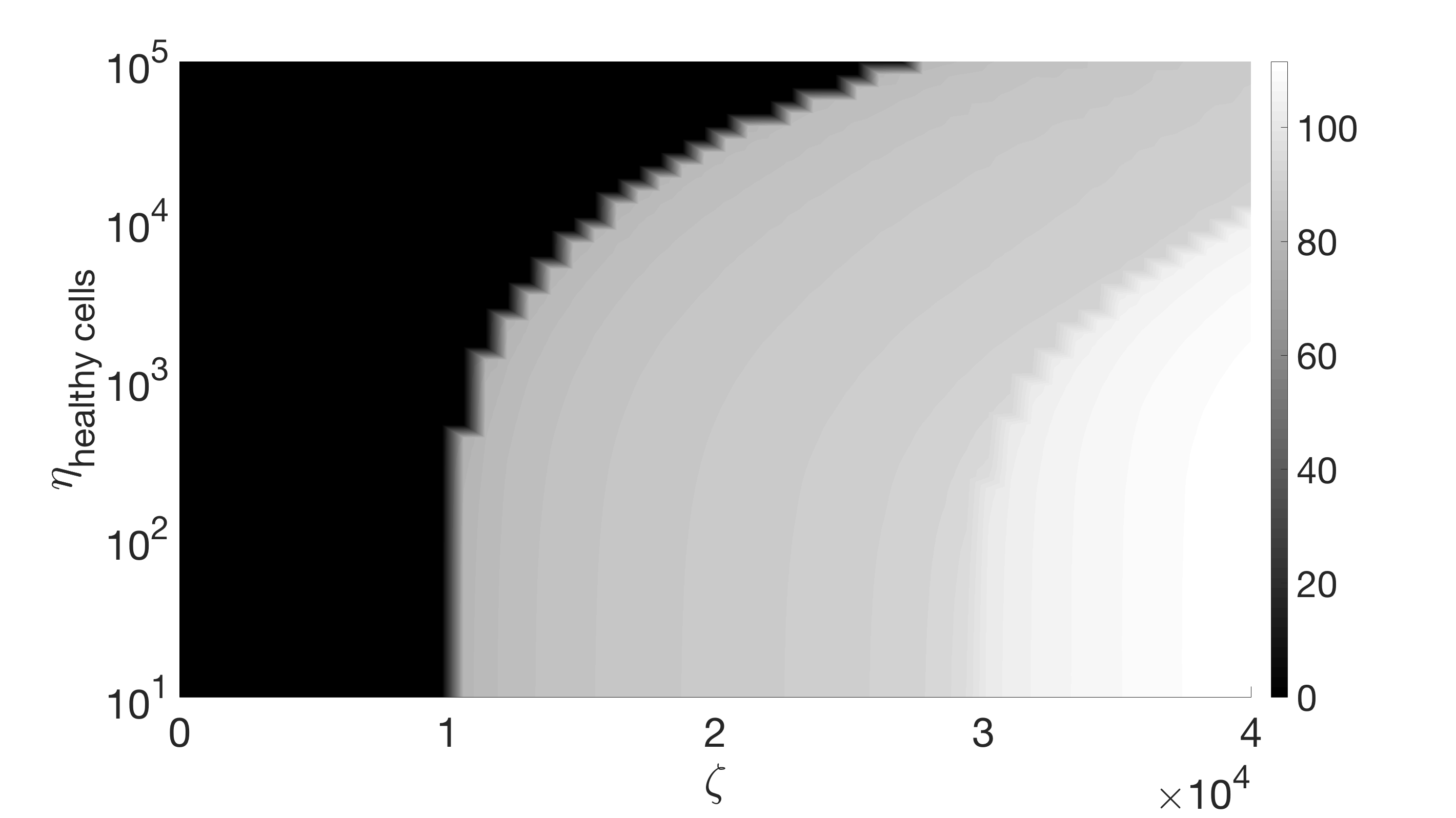}
	\caption{Plot of displacement metric, $d_{\text{max}}$. We calculate this for any $(\eta_{\text{healthy cells}}, \zeta)$ by finding the displacement contour corresponding to $0.1$ voxels and computing the maximum distance of this contour from the location of the initial tumor seed (which is fixed). Given an understanding of how large deformations are typically in glioblastoma growth,  this figure shows us how specific ranges of $(\eta_{\text{healthy cells}}, \zeta)$ can be more useful in obtaining a realistic tissue displacement.} \label{f:mes_b}
\end{figure}

~\fref{f:mes_a} shows that the deformation Jacobian increases with higher forcing factors and smaller screening, which is unsurprising. The localization effect of the screening coefficient is captured in~\fref{f:mes_b}. This results provides us with information regarding the range of screening coefficients and forcing factors that would be useful to capture realistic deformations.


\subsection{Mesh Convergence}
We perform a simple mesh convergence study. We discretize uniformly in all directions using mesh size $h = 2\pi/512$ as the ground truth/reference and report the error measure $e_{\iota}$ for species $\iota$, defined as:
\begin{equation}
e_{\iota} = \frac{||\iota_h - \iota_{\text{ref}}||_2}{||\iota_{\text{ref}}||_2},
\end{equation}
where $\iota_h$ is the concentration of species $\iota$ at any mesh size $h$ and $\iota_{\text{ref}}$ is the reference concentration of species $\iota$. We refine both space and time together to observe our numerical convergence rate. We report our results in \tref{t:mesh_convergence}. We can observe an approximate first order of convergence for our numerical schemes, as expected.

\begin{center}
	\def\arraystretch{1.2}
	\begin{table}[!htbp]
		\caption{$L_2$ convergence error rates $e_{\iota}$ for different model species. The table shows an approximate linear convergence rate as expected for our numerical methods.} 
		\label{t:mesh_convergence}
		\centering
		\begin{tabular}{c|c|c|c|c|c} 
			\hline
			\textbf{Mesh size, $h$} & \textbf{$e_p$} & \textbf{$e_i$} &\textbf{$e_n$} & \textbf{$e_g$} & \textbf{$e_w$}\\  
			\hline
			$2\pi/64$ &   0.5677  &    0.1904  &    0.1677   &   0.2984    &   0.2421 \\
			\ghligh
			$2\pi/128$ &    0.3437 &   0.1483   &    0.0986   &  0.1379    &  0.1171  \\
			$2\pi/256$ &    0.1510 &   0.0645   &     0.0382  &   0.0291   &  0.0292  \\			
			\hline
		\end{tabular}
	\end{table}
\end{center}  

\section{Conclusions and Future Work}\label{s:conclusions}
We presented a model that captures the phenomenological features of glioblastomas seen on MRI scans. These features  include a multicomponent structure of a glioblastoma and tumor-induced mass effect on surrounding brain tissue. We coupled a multispecies ``go-or-grow" tumor model with linear elasticity equations and presented results to illustrate the capabilities of our model in capturing different tumor characteristics using novel numerical schemes. We are currently working on extending the sensitivity analysis and understanding the important parameters of the model by formulating an inverse problem and directly inverting for parameters from images using adjoint-based gradient methods.
\bibliographystyle{apalike} 
\bibliography{refs}
\end{document}